\documentclass[AMA,STIX1COL]{WileyNJD-v2}

\usepackage{amsthm}
\usepackage{mathtools}
\usepackage{tuda-pgfplots}
\usepackage{siunitx}
\usepackage[european]{circuitikz}

\articletype{Original Article}

\received{}
\revised{}
\accepted{}

\newcommand{\mb}[1]{\mathbf{#1}}
\newcommand{\mbt}[1]{\tilde{\mathbf{#1}}}
\newcommand{\mbb}[1]{\bar{\mathbf{#1}}}
\newcommand{\mbh}[1]{\hat{\mathbf{#1}}}
\newcommand{\mr}[1]{\mathrm{#1}}
\newcommand{\T}{{\!\top}}
\newcommand{\ddt}{\frac{\mathrm{d}}{\mathrm{d}t}}
\newcommand{\A}[1]{\mb{A}_\mr{#1}}
\newcommand{\AT}[1]{\mb{A}_\mr{#1}^{\T}}
\newcommand{\qC}{\mb{q}_\mr{C}}
\newcommand{\gR}{\mb{g}_\mr{R}}
\newcommand{\phiL}{\boldsymbol{\phi}_\mr{L}}
\newcommand{\vphi}{\boldsymbol{\varphi}}
\renewcommand{\i}[1]{\mb{i}_\mr{#1}}
\renewcommand{\v}[1]{\mb{v}_\mr{#1}}

\usetikzlibrary{external}

\begin{document}

\title{Index-aware learning of circuits}

\author[1]{Idoia Cortes Garcia}
\author[1,2]{Peter Förster*}
\author[2]{Lennart Jansen}
\author[1]{Wil Schilders}
\author[2]{Sebastian Schöps}
\address[1]{\orgname{Eindhoven University of Technology}, \orgaddress{\country{Netherlands}}}
\address[2]{\orgname{Technical University of Darmstadt}, \orgaddress{\country{Germany}}}
\corres{*Peter Förster, Eindhoven University of Technology. \email{p.f.forster@tue.nl}}

\abstract[Summary]{
    Electrical circuits are present in a variety of technologies, making their design an important part of computer aided engineering. The growing number of parameters that affect the final design leads to a need for new approaches to quantify their impact. Machine learning may play a key role in this regard, however current approaches often make suboptimal use of existing knowledge about the system at hand. In terms of circuits, their description via modified nodal analysis is well-understood. This particular formulation leads to systems of differential-algebraic equations (DAEs) which bring with them a number of peculiarities, e.g. hidden constraints that the solution needs to fulfill. We use the recently introduced dissection index that can decouple a given system of DAEs into ordinary differential equations, only depending on differential variables, and purely algebraic equations, that describe the relations between differential and algebraic variables. The idea is to then only learn the differential variables and reconstruct the algebraic ones using the relations from the decoupling. This approach guarantees that the algebraic constraints are fulfilled up to the accuracy of the nonlinear system solver, and it may also reduce the learning effort as only the differential variables need to be learned.}


\maketitle

\section{Introduction}
\label{sec:int}
Design optimization and uncertainty quantification are key tools of modern computer aided engineering, that both rely on objective functions to express quantities of interest in terms of the variables of the underlying system. Due to the increasing complexity of engineering systems, machine learning approaches have gained popularity for constructing surrogate models of objective functions when they become expensive to evaluate and a large number of (design or uncertainty) parameters are present. In such situations, classical model order reduction techniques\cite{schilders2008} or function approximation approaches\cite{xiu2010} suffer from the curse of dimensionality: the number of operations to construct and the memory required to store the surrogate model grow exponentially with respect to the number of parameters. Experimental and in some cases even theoretical evidence\cite{beneventano2020} shows that machine learning approaches may be able to overcome this curse of dimensionality and provide surrogate models that are fast to evaluate, while requiring comparably little data for their construction and storage.

In the context of electrical circuit design, neural networks have been used for design optimization for over 20 years\cite{zaabab1995,fang2000}. More recently, Gaussian process regression has been employed for both uncertainty quantification and design optimization during analog integrated circuit design\cite{wang2018,sanabria2020}. The commonality between these approaches is that they focus on the learning part: they all aim to provide a computationally efficient and accurate surrogate model, given data produced by some circuit simulator. Thus, they all treat the circuit simulator as a black box that simply provides the data which is then used for constructing the surrogate. In contrast, we want to exploit the known structure that underlies the equations describing electrical circuits.

More specifically, we consider the modified nodal analysis\cite{ho1975} (MNA). MNA is one of the most popular circuit descriptions and lies at the center of SPICE-like simulation software such as LTspice\cite{ltspice}, Xyce\cite{xyce} and PSpice\cite{pspice}. Applying MNA to a given circuit leads to systems of differential-algebraic equations (DAEs), which can generally be written as systems of implicit differential equations \cite{denk1991}
\begin{align}
    \mb{F}(\mb{x}', \mb{x}, t, \mb{p}) = \mb{0}, \quad \mb{x}(0) = \mb{x}_0, \label{eq:int_dae}
\end{align}
where the Jacobian $\mb{J}_{\mb{x}'}(\mb{F})$, of $\mb{F}$ w.r.t.~$\mb{x}'$, is singular and $\mb{p}$ are the design or uncertainty parameters. Intuitively, one can think of DAEs as ordinary differential equations (ODEs) that are constrained to a manifold determined by (hidden) constraints on the solution variables $\mb{x}$. We aim to exploit the special structure of the DAEs arising from MNA by using the dissection index\cite{jansen2014} to propose an approach for learning electrical circuits more accurately and efficiently. More concretely, we use the dissection index to decouple the DAEs into sets of ODEs and purely algebraic equations, such that the entire dynamics of the solution may then be found only using the ODEs, while the algebraic equations may be used to recover the entire solution.

In the following, \autoref{sec:md} introduces MNA and DAEs in more detail and states some well-known results. Afterwards, \autoref{sec:di} outlines the dissection index and showcases its properties using example circuits. The new approach is then presented on an abstract level in \autoref{sec:ial}, and on a numerical level in \autoref{sec:ne}. Preliminary conclusions about the effectiveness of the approach and future research directions are given in \autoref{sec:cfr}.

\section{MNA and DAEs}
\label{sec:md}
We first look at the system of DAEs that results from MNA when not considering controlled sources. As they are of crucial importance for engineering applications however, we will note whether extensions including controlled sources are available or missing at the appropriate times. Borrowing the notation from Tischendorf\cite{tischendorf1999}, the system of MNA reads
\begin{subequations}
    \label{eq:md_mnai}
    \begin{align}
            \A{C} \ddt \qC \big( \AT{C} \vphi \big) + \A{R} \gR \big( \AT{R} \vphi \big) + \A{L} \i{L} + \A{V} \i{V} + \A{I} \i{s}(t) &= \mb{0}\\ 
            \ddt \phiL \big( \i{L} \big) - \AT{L} \vphi &= \mb{0}\\
            \AT{V} \vphi - \v{s}(t)&= \mb{0},
    \end{align}
\end{subequations}
where the left hand side as a whole corresponds to $\mb{F}$ in \eqref{eq:int_dae}, and the solution variables are given by $\mb{x} = [\vphi, \i{L}, \i{V}]^\T$. The functions $\gR$, $\phiL$ and $\qC$ model resistive, capacitive or inductive devices respectively, that may each depend on the parameters $\mb{p}$. The terms for independent current and voltage sources are given by $\i{s}(t)$ and $\v{s}(t)$, while $\vphi$ denotes the vector of nodal potentials and $\i{L} $, $\i{V}$ are the currents flowing through branches containing inductors or voltage sources respectively. The last ingredient is given by the incidence matrices $\A{*}$, where $*$ indicates the device type. These collect the branch to node relations of the underlying electrical network, when considering the branches and nodes as edges and vertices of a directed graph.

In order to obtain a version of \eqref{eq:md_mnai} that is better suited to analysis and implementation, we consider the device function Jacobians\footnote{Note that the definition of $\mb{G}$ via the Jacobian only serves to obtain a matrix form in \eqref{eq:md_mnaii} and has no impact on the general approach. As such, one may also work directly with the original system from \eqref{eq:md_mnai}, however software implementations of MNA often work with $\mb{G}$ as defined in \eqref{eq:md_dfj} (e.g.~when using Newton's method for solving nonlinear systems).}
\begin{align}
    \mb{G} \big( \AT{R} \vphi \big) \coloneqq \mb{J}_{\AT{R} \vphi}(\gR), \quad \mb{L}(\i{L}) \coloneqq \mb{J}_{\i{L}}(\phiL), \quad \mb{C} \big( \AT{C} \vphi \big) \coloneqq \mb{J}_{\AT{C} \vphi}(\qC), \label{eq:md_dfj}
\end{align}
where we use the same notation for the Jacobians as in \autoref{sec:int}. Inserting \eqref{eq:md_dfj} into the original system and writing everything in matrix form yields
\begin{align}
    \begin{bmatrix}
        \A{C}^{\phantom{\T}} \mb{C} \big( \AT{C} \vphi \big) \AT{C} & &\\
        & \mb{L}(\i{L}) &\\
        & & \mb{0}
    \end{bmatrix} \ddt \begin{bmatrix}
        \vphi\\
        \i{L}\\
        \i{V}
    \end{bmatrix} + \begin{bmatrix}
        \A{R}^{\phantom{\T}} \mb{G} \big( \AT{R} \vphi \big) \AT{R} & \A{L} & \A{V}\\
        -\AT{L} & &\\
        -\AT{V} & &
    \end{bmatrix} \begin{bmatrix}
        \vphi\\
        \i{L}\\
        \i{V}
    \end{bmatrix} + \begin{bmatrix}
        \A{I} \i{s}(t)\\
        \mb{0}\\
        \v{s}(t)
    \end{bmatrix} = \mb{0}, \label{eq:md_mnaii}
\end{align}
assuming that $\phiL$ and $\qC$ do not explicitly depend on time.

We note that \eqref{eq:md_mnaii} readily extends to multiport devices\cite{tischendorf1999}. A general inductive $n$-port for example, can also be modeled by a function $\phiL(\i{L}) = \big[ \phi_1(\i{L}), \dotsc, \phi_{n-1}(\i{L}) \big]^\T$, however the Jacobian is not necessarily diagonal in this case as the component functions of $\phiL$ can each depend on all the currents $\i{L} = [i_1, \cdots, i_{n-1}]^\T$. Regarding the incidence matrices, one chooses a reference node $\varphi_0$ for the multiport device and then considers $n-1$ branches from the remaining $n-1$ nodes to the reference. Interpreting the multiport as a single node and using Kirchhoff's current law then gives $i_0 = -\sum_{k=1}^{n-1} i_k$ for the reference current, when orienting all currents to point toward the device. Following this approach, there is then no difference in treating multiports compared to one-ports\cite{tischendorf1999}.

\subsection{Index concepts}
\label{subsec:ic}
Before outlining the dissection index, we want to give a brief introduction to index concepts more generally. There are multiple definitions of the index of a DAE, along with related index concepts that each possess different strengths and weaknesses\cite{mehrmann2015}. One important aspect that unites these ideas is that they agree in key cases, e.g.~when looking at linear DAEs, and the same also holds true for the dissection index. To emphasize the practical importance of the notion of index, we take a closer look at the perturbation index. It is based on a perturbed version of the DAE \eqref{eq:int_dae} (we leave out the parameters $\mb{p}$ for conciseness)
\begin{align}
    \mb{F}(\mbh{x}', \mbh{x}, t) = \boldsymbol{\varepsilon}(t), \quad \mbh{x}(0) = \mbh{x}_0, \label{eq:ic_pdae}
\end{align}
where $\varepsilon$ is a sufficiently smooth perturbation, such that the required derivatives exist.
\begin{definition}
    \label{def:ic_pi}
    Let $\mb{x}$ be a solution of the unperturbed DAE, then the DAE is said to have perturbation index $\nu \in \mathbb{N}$, if $\nu$ is the smallest natural number such that, for any sufficiently smooth solution $\mbh{x}$ of \eqref{eq:ic_pdae}, there exists $c \in \mathbb{R}$ with
    \begin{align*}
        \| \mbh{x} - \mb{x} \| \leq c \big( \| \mbh{x}_0 - \mb{x}_0 \| + \| \boldsymbol{\varepsilon} \|_\infty + \| \boldsymbol{\varepsilon}' \|_\infty + \dotsb + \| \boldsymbol{\varepsilon}^{(\nu - 1)} \|_\infty \big)
    \end{align*}
    for an appropriate norm $\| \cdot \|$ and the right hand side small enough.
\end{definition}
The idea behind this definition is to capture the impact of perturbations on the solution, as the name suggests. Usually these pertubations are assumed small, in the sense that $\| \boldsymbol{\varepsilon} \|_\infty \ll 1$, but fast changing, such that $\| \boldsymbol{\varepsilon}^{(\nu)} \|_\infty$ may grow very quickly in $\nu$.

\paragraph{Example}
\begin{figure}[b]
    \begin{center}
        \includegraphics[width=0.35\textwidth]{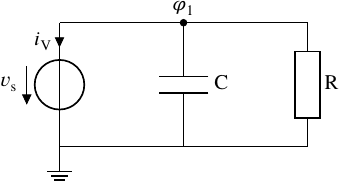}
    \end{center}
    \caption{Small example circuit for illustrating the perturbation index.}
    \label{fig:ic_vcr}
\end{figure}
To illustrate the perturbation index, and also to hint at its relevance for circuit simulation, we consider the small example given in \autoref{fig:ic_vcr}. Applying MNA to the circuit and introducing a perturbation yields the following DAE
\begin{subequations}
    \label{eq:ic_vcr}
    \begin{align}
        C \ddt \hat{\varphi}_1 + \frac{1}{R} \hat{\varphi}_1 + \hat{i}_\mr{V} &= \varepsilon_1(t) \label{eq:ic_vcra}\\
        -\hat{\varphi}_1 + v_\mr{s}(t) &= \varepsilon_2(t), \label{eq:ic_vcrb}
    \end{align}
\end{subequations}
where $\mbh{x} = [\hat{\varphi}_1, \hat{i}_\mr{V}]^\T$. Using \eqref{eq:ic_vcrb} we find
\begin{align}
    \hat{\varphi}_1 = v_\mr{s}(t) - \varepsilon_2(t). \label{eq:ic_vcrphi}
\end{align}
Inserting \eqref{eq:ic_vcrphi} into \eqref{eq:ic_vcra} and rearranging then gives
\begin{align*}
    \hat{i}_\mr{V} &= \varepsilon_1(t) - C \ddt \big( v_\mr{s}(t) - \varepsilon_2(t) \big) - \frac{1}{R} \big( v_\mr{s}(t) - \varepsilon_2(t) \big).
\end{align*}
Noting that the solution $\mb{x}$ to the unperturbed problem follows directly from the perturbed solution by setting $\boldsymbol{\varepsilon} = \mb{0}$, we obtain
\begin{align}
    \| \mbh{x} - \mb{x} \| = \begin{Vmatrix}
        -\varepsilon_2(t)\\
        \varepsilon_1(t) + C \ddt \varepsilon_2(t) + \frac{1}{R} \varepsilon_2(t)
    \end{Vmatrix}, \label{eq:ic_vcrpi}
\end{align}
so the unperturbed system corresponding to \eqref{eq:ic_vcr} has perturbation index $\nu = 2$ as \eqref{eq:ic_vcrpi} depends on the first derivative of $\varepsilon_2$.

\paragraph{Index of MNA}
The structure and index of MNA are well understood when only considering independent sources\cite{tischendorf1999} (as in \eqref{eq:md_mnaii}), but also when including controlled sources\cite{estevez2000ijcta}. For the case without controlled sources, there exists the following well-known topological index result.
\begin{theorem}
    \label{th:ic_imna}
    Assuming the matrix-valued functions $\mb{G} \big( \AT{R} \vphi \big)$, $\mb{L}(\i{L})$, $\mb{C} \big( \AT{C} \vphi \big)$ from \eqref{eq:md_dfj} positive definite:
    \begin{enumerate}
        \item The index of \eqref{eq:md_mnaii} is $\nu \leq 2$.
        \item The index of \eqref{eq:md_mnaii} is $\nu \leq 1$, if and only if there are no loops consisting only of capacitors and voltage sources containing at least one voltage source and no cutsets consisting only of inductors and current sources.
    \end{enumerate}
\end{theorem}
The result can be found in terms of the perturbation index\cite{gunther1996}, the tractability index\cite{tischendorf1999}, the differentiation index\cite{estevez2000ijcta} and the dissection index\cite{jansen2014}. We again note that there also exist extensive results about when the index of MNA including controlled sources does not exceed $\nu = 2$\cite{estevez2000ijcta}.

\section{Dissection index}
\label{sec:di}
We focus on the dissection index, as it enables the decoupling of a DAE into an ODE and a set of purely algebraic equations. This is conceptually different from the perturbation index, however other index concepts such as the tractability and differentiation indices also use decoupling strategies. Still, the dissection index maintains some advantages over these concepts, as it provides a simple algorithmic procedure for the decoupling that is similar to the tractability index, but poses less strict smoothness assumptions. In the case of MNA without controlled sources \eqref{eq:md_mnaii}, it is even possible to find a purely topological decoupling based on the dissection index\cite{jansen2014}. We will not make use of this topological decoupling in the derivation however, but rather consider the dissection index for a more general class of DAEs, to formulate the assumptions that are necessary for our method to work also for DAEs other than \eqref{eq:md_mnaii}.

We consider a DAE in standard form\cite{jansen2014},
\begin{align}
    \mb{M}(\mb{x}) \ddt \mb{x} + \mb{K}(\mb{x}) \mb{x} + \mb{f}(t) = \mb{0}, \quad \ker \mb{M} \supsetneq \{ \mb{0} \}, \label{eq:di_dae}
\end{align}
where the matrix-valued functions $\mb{M}$ and $\mb{K}$ are derived from \eqref{eq:int_dae} by defining
\begin{align*}
    \mb{M}(\mb{x}) = \mb{J}_{\mb{x}'} \big( \mb{F}(\mb{x}', \mb{x}, t) \big), \quad \mb{K}(\mb{x}) = \mb{J}_{\mb{x}} \big( \mb{F}(\mb{x}', \mb{x}, t) \big).
\end{align*}
Note that the description of MNA in \eqref{eq:md_mnaii} is precisely of this form. In the following, we will demonstrate the first two steps of the dissection index when applied to systems of the form of \eqref{eq:di_dae}, while stating the assumptions of our approach. Appendix \autoref{sec:dim} contains additional remarks showing that these assumptions are fulfilled by \eqref{eq:md_mnaii}.

\subsection{Index one case}
\label{subsec:ioc}
Assuming $\mb{M}(\mb{x})$ to be sufficiently smooth with values in $\mathbb{R}^{n \times m}$, we define four basis functions $\mb{P}(\mb{x})$, $\mb{Q}(\mb{x})$, $\mb{V}(\mb{x})$, $\mb{W}(\mb{x})$ such that
\begin{align*}
    \mr{im\, } \mb{Q}(\mb{x}) = \ker \mb{M}(\mb{x}), \quad \mr{im\, } \mb{W}(\mb{x}) = \ker \mb{M}^\T (\mb{x})
\end{align*}
and the columns of $\mb{P}(\mb{x})$ and $\mb{Q}(\mb{x})$ together form a basis of $\mathbb{R}^n$, while the columns of $\mb{V}(\mb{x})$ and $\mb{W}(\mb{x})$ together form a basis of $\mathbb{R}^m$. We now make the following assumption.
\begin{assumption}
    \label{as:ioc_bf}
    The basis functions $\mb{P}$ and $\mb{Q}$ of $\mb{M}(\mb{x})$ are constant.
\end{assumption}
\autoref{as:ioc_bf} may seem restrictive, but it is fulfilled by many systems occurring in practical applications\cite{jansen2014}. This in particular includes MNA, as \autoref{re:dim_bf} shows. The key idea of the dissection index is to use the basis functions to split the solution variables $\mb{x}$ into two parts
\begin{align}
    \mb{x} = \mb{P} \mbt{x} + \mb{Q} \mbb{x}, \label{eq:ioc_x}
\end{align}
where $\tilde{\cdot}$ is used to indicate differential (dynamic) variables and $\bar{\cdot}$ signifies algebraic (fixed) variables. When inserting the splitting \eqref{eq:ioc_x} into \eqref{eq:di_dae} we obtain
\begin{align}
    \mb{M}(\mb{x}) \mb{P} \ddt \mbt{x} + \mb{K}(\mb{x}) \mb{P} \mbt{x} + \mb{K}(\mb{x}) \mb{Q} \mbb{x} + \mb{f}(t) = \mb{0} \label{eq:ioc_dae}
\end{align}
and this motivates the notation, as only $\mbt{x}$ appears differentiated in time. The procedure then continues by multiplying \eqref{eq:ioc_dae} once with $\mb{V}^\T (\mb{x})$ and $\mb{W}^\T (\mb{x})$ each from the left, to also split the system. This yields
\begin{subequations}
    \label{eq:ioc_io}
    \begin{align}
            \underbrace{\mb{V}^\T (\mb{x}) \mb{M}(\mb{x}) \mb{P}}_{\eqqcolon \mbt{M}(\mb{x})} \ddt \mbt{x} + \underbrace{\mb{V}^\T (\mb{x}) \mb{K}(\mb{x}) \mb{P}}_{\eqqcolon \mbt{K}_\mb{P}(\mb{x})} \mbt{x} + \underbrace{\mb{V}^\T (\mb{x}) \mb{K}(\mb{x}) \mb{Q}}_{\eqqcolon \mbt{K}_\mb{Q}(\mb{x})} \mbb{x} + \underbrace{\mb{V}^\T (\mb{x}) \mb{f}(t)}_{\eqqcolon \mbt{f}(t)} &= \mb{0} \label{eq:ioc_ioa}\\
            \underbrace{\mb{W}^\T (\mb{x}) \mb{K}(\mb{x}) \mb{P}}_{\eqqcolon \mbb{K}_\mb{P}(\mb{x})} \mbt{x} + \underbrace{\mb{W}^\T (\mb{x}) \mb{K}(\mb{x}) \mb{Q}}_{\eqqcolon \mbb{K}_\mb{Q}(\mb{x})} \mbb{x} + \underbrace{\mb{W}^\T (\mb{x}) \mb{f}(t)}_{\eqqcolon \mbb{f}(t)} &= \mb{0}, \label{eq:ioc_iob}
    \end{align}
\end{subequations}
where we also introduce shorthands for the arising products. Using the fact that $\mbt{M}(\mb{x})$ is regular by construction\cite{jansen2014}, we can now define the index one case of the dissection index.
\begin{definition}
    \label{def:ioc_io}
    The DAE \eqref{eq:di_dae} has dissection index $\nu = 1$ if $\mbb{K}_\mb{Q}(\mb{x})$ is regular.
\end{definition}
This is motivated by the observation that \eqref{eq:ioc_iob} is a purely algebraic equation with a locally unique solution for $\mbb{x}$ in terms of $\mbt{x}$, given that $\mbb{K}_\mb{Q}(\mb{x})$ is regular. In this case \eqref{eq:ioc_ioa} then describes an ODE in the differential variables $\mbt{x}$.

\subsection{Index two case}
\label{subsec:itc}
As the example from \autoref{fig:ic_vcr} illustrates, there are many DAEs, including those described by MNA, which can have an index higher than one. In these cases the dissection index proceeds by introducing additional basis functions and continuing the splitting process in a similar fashion. We begin by focusing on the algebraic equation \eqref{eq:ioc_iob}, and consider basis functions $\mbb{P}(\mb{x})$, $\mbb{Q}(\mb{x})$, $\mbb{V}(\mb{x})$, $\mbb{W}(\mb{x})$ of $\mbb{K}_\mb{Q}(\mb{x})$, defined analogous to the ones for $\mb{M}(\mb{x})$. This allows us to further split the algebraic variables $\mbb{x}$ as follows
\begin{align}
    \mbb{x} = \mbb{P}(\mb{x}) \mbb{x}_\mb{P} + \mbb{Q}(\mb{x}) \mbb{x}_\mb{Q}. \label{eq:itc_xb}
\end{align}
Inserting this splitting into \eqref{eq:ioc_iob}, and multiplying once by $\mbb{V}^\T (\mb{x})$ and $\mbb{W}^\T (\mb{x})$ each from the left, splits the algebraic equation into two parts
\begin{subequations}
    \label{eq:itc_iti}
    \begin{align}
        \mbb{V}^\T (\mb{x}) \mbb{K}_\mb{P}(\mb{x}) \mbt{x} + \mbb{V}^\T (\mb{x}) \mbb{K}_\mb{Q}(\mb{x}) \mbb{P}(\mb{x}) \mbb{x}_\mb{P} + \mbb{V}^\T (\mb{x}) \mbb{f}(t) &= \mb{0} \label{eq:itc_itia}\\
        \mbb{W}^\T (\mb{x}) \mbb{K}_\mb{P}(\mb{x}) \mbt{x} + \mbb{W}^\T (\mb{x}) \mbb{f}(t) &= \mb{0}. \label{eq:itc_itib}
    \end{align}
\end{subequations}
We can now also split the differential variables $\mbt{x}$ further, by using basis functions $\mbt{P}(\mb{x})$ and $\mbt{Q}(\mb{x})$ of $\mbb{W}^\T (\mb{x}) \mbb{K}_\mb{P}(\mb{x})$
\begin{align}
    \mbt{x} = \mbt{P}(\mb{x}) \mbt{x}_\mb{P} +
    \mbt{Q}(\mb{x}) \mbt{x}_\mb{Q}, \label{eq:itc_xt}
\end{align}
and inserting this splitting into the second algebraic equation \eqref{eq:itc_itib} yields
\begin{align}
    \mbb{W}^\T (\mb{x}) \mbb{K}_\mb{P}(\mb{x}) \mbt{P}(\mb{x}) \mbt{x}_\mb{P} + \mbb{W}^\T (\mb{x}) \mbb{f}(t) &= \mb{0}. \label{eq:itc_xtp}
\end{align}
We now make the following two assumptions.
\begin{assumption}
    \label{as:itc_xtpdi}
    The matrix product $\mbb{W}^\T (\mb{x}) \mbb{K}_\mb{P}(\mb{x}) \mbt{P}(\mb{x}) \mbt{x}_\mb{P}$ is regular.
\end{assumption}
\begin{assumption}
    \label{as:itc_xtp}
    The basis functions $\mbt{P}$ and $\mbt{Q}$ of $\mbb{W}^\T (\mb{x}) \mbb{K}_\mb{P}(\mb{x})$ are constant and \eqref{eq:itc_xtp} possesses a locally unique solution for $\mbt{x}_\mb{P}$ in terms of $\mbt{x}_\mb{Q}$ and $t$.
\end{assumption}
We note that while \autoref{as:itc_xtpdi} is equivalent to the DAE not being underdetermined\cite{jansen2014}, \autoref{as:itc_xtp} is rather important for the implementation, but not for the dissection index itself. In fact our approach still works if the basis functions $\mbt{P}(\mbt{x}_\mb{Q})$ and $\mbt{Q}(\mbt{x}_\mb{Q})$ depend on $\mbt{x}_\mb{Q}$, however we focus on the stronger assumption here, since it shortens the expressions in the following without impacting the general idea and \autoref{re:dim_xtp} shows that MNA fulfills an even stronger condition than \autoref{as:itc_xtp}.

Having a locally unique solution for $\mbt{x}_\mb{P}$ in terms of $\mbt{x}_\mb{Q}$ and $t$ at hand, we now turn to the first algebraic equation \eqref{eq:itc_itia}. Similar to \eqref{eq:itc_xtp}, inserting the splitting \eqref{eq:itc_xt} of the differential variables into \eqref{eq:itc_itia} gives a system with a locally unique solution for $\mbb{x}_\mb{P}$ in terms of $\mbt{x}_\mb{Q}$, $\mbt{x}_\mb{P}$, $\mbb{x}_\mb{Q}$ and $t$, as $\mbb{V}^\T (\mb{x}) \mbb{K}_\mb{Q}(\mb{x}) \mbb{P}(\mb{x})$ is regular by construction\cite{jansen2014}
\begin{align}
    \mbb{V}^\T (\mb{x}) \mbb{K}_\mb{P}(\mb{x}) \mbt{P} \mbt{x}_\mb{P} + \mbb{V}^\T (\mb{x}) \mbb{K}_\mb{P}(\mb{x}) \mbt{Q} \mbt{x}_\mb{Q} + \mbb{V}^\T (\mb{x}) \mbb{K}_\mb{Q}(\mb{x}) \mbb{P}(\mb{x}) \mbb{x}_\mb{P} + \mbb{V}^\T (\mb{x}) \mbb{f}(t) &= \mb{0}. \label{eq:itc_xbp}
\end{align}
Finally, we move towards the decoupled index two system by expanding $\mbt{x}$ and $\mbb{x}$ in \eqref{eq:ioc_ioa}, according to \eqref{eq:itc_xt} and \eqref{eq:itc_xb} respectively,
\begin{align}
    \mbt{M}(\mb{x}) \mbt{P} \ddt \mbt{x}_\mb{P} + \mbt{M}(\mb{x}) \mbt{Q} \ddt \mbt{x}_\mb{Q} + \mbt{K}_\mb{P}(\mb{x}) \mbt{P} \mbt{x}_\mb{P} + \mbt{K}_\mb{P}(\mb{x}) \mbt{Q} \mbt{x}_\mb{Q} + \mbt{K}_\mb{Q}(\mb{x}) \mbb{P}(\mb{x}) \mbb{x}_\mb{P} + \mbt{K}_\mb{Q}(\mb{x}) \mbb{Q}(\mb{x}) \mbb{x}_\mb{Q} + \mbt{f}(t) &= \mb{0}. \label{eq:itc_itii}
\end{align}
Using basis functions $\mbt{V}(\mb{x})$ and $\mbt{W}(\mb{x})$ of $\mbt{M}(\mb{x}) \mbt{Q}$, we can now split \eqref{eq:itc_itii} further by multiplying from the left by $\mbt{V}^\T (\mb{x})$ and $\mbt{W}^\T (\mb{x})$ once each. Reordering then yields
\begin{subequations}
    \label{eq:itc_itiii}
    \begin{align}
            \mbt{V}^\T (\mb{x}) \mbt{M}(\mb{x}) \mbt{Q} \ddt \mbt{x}_\mb{Q} + \mbt{V}^\T (\mb{x}) \mbt{K}_\mb{P}(\mb{x}) \mbt{Q} \mbt{x}_\mb{Q} + \mbt{V}^\T (\mb{x}) \mbt{K}_\mb{Q}(\mb{x}) \mbb{Q}(\mb{x}) \mbb{x}_\mb{Q} + \mbt{V}^\T (\mb{x}) \big( \mbt{M}(\mb{x}) \mbt{P} \ddt \mbt{x}_\mb{P} + \mbt{K}_\mb{P}(\mb{x}) \mbt{P} \mbt{x}_\mb{P} + \mbt{K}_\mb{Q}(\mb{x}) \mbb{P}(\mb{x}) \mbb{x}_\mb{P} + \mbt{f}(t) \big) &= \mb{0} \label{eq:itc_itiiia}\\
            \mbt{W}^\T (\mb{x}) \mbt{K}_\mb{P}(\mb{x}) \mbt{Q} \mbt{x}_\mb{Q} + \mbt{W}^\T (\mb{x}) \mbt{K}_\mb{Q}(\mb{x}) \mbb{Q}(\mb{x}) \mbb{x}_\mb{Q} + \mbt{W}^\T (\mb{x}) \big( \mbt{M}(\mb{x}) \mbt{P} \ddt \mbt{x}_\mb{P} + \mbt{K}_\mb{P}(\mb{x}) \mbt{P} \mbt{x}_\mb{P} + \mbt{K}_\mb{Q}(\mb{x}) \mbb{P}(\mb{x}) \mbb{x}_\mb{P} + \mbt{f}(t) \big) &= \mb{0}. \label{eq:itc_itiiib}
    \end{align}
\end{subequations}
We observe that \eqref{eq:itc_itiii} is of a form similar to \eqref{eq:ioc_io} and that $\mbt{V}^\T (\mb{x}) \mbt{M}(\mb{x}) \mbt{Q}$ is again regular by construction\cite{jansen2014}. Together with \eqref{eq:itc_xtp} and \eqref{eq:itc_xbp} providing locally unique solutions for $\mbt{x}_\mb{P}$ and $\mbb{x}_\mb{P}$ respectively, this motivates the following definition.
\begin{definition}
    \label{def:itc_it}
     The DAE has dissection index $\nu = 2$ if $\mbt{W}^\T (\mb{x}) \mbt{K}_\mb{Q}(\mb{x}) \mbb{Q}(\mb{x})$ is regular.
\end{definition}
This again follows from the purely algebraic equation \eqref{eq:itc_itiiib} having a locally unique solution for $\mbb{x}_\mb{Q}$ in terms of $\mbt{x}_\mb{Q}$, $\mbt{x}_\mb{P}$, $\mbb{x}_\mb{P}$ and $t$, given that $\mbt{W}^\T (\mb{x}) \mbt{K}_\mb{Q}(\mb{x}) \mbb{Q}(\mb{x})$ is regular. The differential part \eqref{eq:itc_itiiia} then describes an ODE in the index two differential variables $\mbt{x}_\mb{Q}$, analogous to the previous case, and we call $\mbt{x}_\mb{Q}$ the differential variables and $[\mbt{x}_\mb{P}, \mbb{x}_\mb{P}, \mbb{x}_\mb{Q}]^\T$ the algebraic variables. We note that the procedure may be continued for even higher index DAEs, by repeating the steps of the index two case with $\mbt{x}_\mb{Q}$ and $\mbb{x}_\mb{Q}$ playing the roles of $\mbt{x}$ and $\mbb{x}$ respectively.

\paragraph{First example circuit}
\begin{figure}[b]
    \begin{center}
        \includegraphics[width=0.5\textwidth]{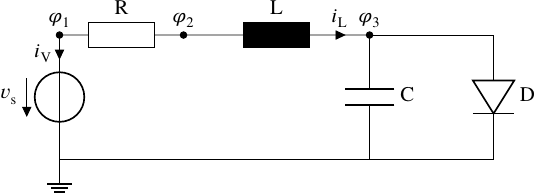}
    \end{center}
    \caption{First example circuit: simple diode oscillator.}
    \label{fig:itc_do1}
\end{figure}
We now demonstrate the dissection index by applying it to the example circuit given in \autoref{fig:itc_do1}. The circuit contains a voltage source $v_\mr{s}(t)$, a linear resistor with resistance $R$, a linear capacitor with capacitance $C$, a linear inductor with inductance $L$, as well as a diode $\mr{D}$ that is modeled by a nonlinear resistance $g_\mr{D}(\varphi_3)$. Comparing the conditions of \autoref{th:ic_imna} with the example circuit shows that the circuit has index $\nu = 1$, thus we only have to perform the first step of the dissection index.

We begin by writing out the system obtained from applying MNA to the example
\begin{align*}
    \begin{bmatrix}
        0 & & & &\\
        & 0 & & &\\
        & & C & &\\
        & & & L &\\
        & & & & 0
    \end{bmatrix} \ddt \begin{bmatrix}
        \varphi_1\\
        \varphi_2\\
        \varphi_3\\
        i_\mr{L}\\
        i_\mr{V}
    \end{bmatrix} + \begin{bmatrix}
        G & -G & &  & 1\\
        -G & G & & 1 &\\
        & & g_\mr{D}(\varphi_3) & -1 &\\
        & -1 & 1 & &\\
        -1 & & & &
    \end{bmatrix} \begin{bmatrix}
        \varphi_1\\
        \varphi_2\\
        \varphi_3\\
        i_\mr{L}\\
        i_\mr{V}
    \end{bmatrix} + \begin{bmatrix}
        0\\
        0\\
        0\\
        0\\
        v_\mr{s}(t)
    \end{bmatrix} = \mb{0},
\end{align*}
where $G = 1/R$ is the inverse of the resistance. For the first basis functions $\mb{Q}$ and $\mb{P}$ we find
\begin{align*}
    \mb{Q} = \begin{bmatrix}
        1 & &\\
        & 1 &\\
        & 0 &\\
        & & 0\\
        & & 1
    \end{bmatrix}, \quad \mb{P} = \begin{bmatrix}
        0 &\\
        0 &\\
        1 &\\
        & 1\\
        & 0
    \end{bmatrix}
\end{align*}
and as $\mb{M}(\mb{x})$ in the context of \eqref{eq:di_dae} is symmetric in this case, we have $\mb{W} = \mb{Q}$ and $\mb{V} = \mb{P}$ for the remaining two basis functions. Using these, we split the unknowns into $\mbt{x} = [\varphi_3, i_\mr{L}]^\T$ and $\mbb{x} = [\varphi_1, \varphi_2, i_\mr{V}]^\T$ according to \eqref{eq:ioc_x}, which allows us to obtain the systems corresponding to \eqref{eq:ioc_ioa} and \eqref{eq:ioc_iob}
\begin{subequations}
    \label{eq:itc_ei}
    \begin{align}
        \begin{bmatrix}
            C &\\
            & L
        \end{bmatrix} \ddt \begin{bmatrix}
            \varphi_3\\
            i_\mr{L}
        \end{bmatrix} + \begin{bmatrix}
            g_\mr{D}(\varphi_3) & -1\\
            1 &
        \end{bmatrix} \begin{bmatrix}
            \varphi_3\\
            i_\mr{L}
        \end{bmatrix} + \begin{bmatrix}
            0 & &\\
            & -1 & 0
        \end{bmatrix} \begin{bmatrix}
            \varphi_1\\
            \varphi_2\\
            i_\mr{V}
        \end{bmatrix} &= \mb{0} \label{eq:itc_eia}\\
        \begin{bmatrix}
            0 &\\
            & 1\\
            & 0
        \end{bmatrix} \begin{bmatrix}
            \varphi_3\\
            i_\mr{L}
        \end{bmatrix} + \begin{bmatrix}
            G & -G & 1\\
            -G & G &\\
            -1 & &
        \end{bmatrix} \begin{bmatrix}
            \varphi_1\\
            \varphi_2\\
            i_\mr{V}
        \end{bmatrix} + \begin{bmatrix}
            0\\
            0\\
            v_\mr{s}(t)
        \end{bmatrix} &= \mb{0}. \label{eq:itc_eib}
    \end{align}
\end{subequations}
Since \eqref{eq:itc_eib} is linear, we can explicitly solve for $\mbb{x}$ in this case. Subsequently inserting this solution into \eqref{eq:itc_eia} then gives an ODE and a purely algebraic system as promised
\begin{align}
    \begin{bmatrix}
        C &\\
        & L
    \end{bmatrix} \ddt \begin{bmatrix}
        \varphi_3\\
        i_L
    \end{bmatrix} + \begin{bmatrix}
        g_\mr{D}(\varphi_3) & -1\\
        1 & R
    \end{bmatrix} \begin{bmatrix}
        \varphi_3\\
        i_\mr{L}
    \end{bmatrix} + \begin{bmatrix}
        0\\
        -v_\mr{s}(t)
    \end{bmatrix} = \mb{0}, \quad \begin{bmatrix}
        \varphi_1\\
        \varphi_2\\
        i_\mr{V}
    \end{bmatrix} = \begin{bmatrix}
        v_\mr{s}(t)\\
        v_\mr{s}(t) - R i_\mr{L}\\
        -i_\mr{L}
    \end{bmatrix}. \label{eq:itc_do1}
\end{align}

\paragraph{Second example circuit}
\begin{figure}[b]
    \begin{center}
        \includegraphics[width=0.5\textwidth]{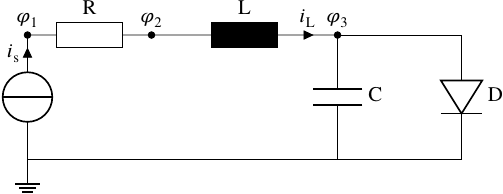}
    \end{center}
    \caption{Second example circuit: simple diode oscillator with current instead of voltage source.}
    \label{fig:itc_do2}
\end{figure}
We derive a second example circuit from the first by substituting a current source $i_\mr{s}(t)$ for the voltage source, compare \autoref{fig:itc_do2}. Looking at the index criteria from \autoref{th:ic_imna}, we observe that this circuit has index $\nu = 2$, as there now is a cutset consisting of the inductor and current source. Therefore, we need to execute two steps of the dissection index in order to split the equations into purely differential and algebraic parts respectively.

The corresponding MNA system is given by
\begin{align*}
    \begin{bmatrix}
        0 & & &\\
        & 0 & &\\
        & & C &\\
        & & & L
    \end{bmatrix} \ddt \begin{bmatrix}
        \varphi_1\\
        \varphi_2\\
        \varphi_3\\
        i_\mr{L}
    \end{bmatrix} + \begin{bmatrix}
        G & -G & &\\
        -G & G & & 1\\
        & & g_\mr{D}(\varphi_3) & -1\\
        & -1 & 1 &
    \end{bmatrix} \begin{bmatrix}
        \varphi_1\\
        \varphi_2\\
        \varphi_3\\
        i_\mr{L}
    \end{bmatrix} + \begin{bmatrix}
        -i_\mr{s}(t)\\
        0\\
        0\\
        0
    \end{bmatrix} = \mb{0},
\end{align*}
where we also note that the dimension is smaller compared to the previous example. The first two basis functions $\mb{Q}$ and $\mb{P}$ are
\begin{align*}
    \mb{Q} = \begin{bmatrix}
        1 &\\
        & 1\\
        & 0\\
        & 0
    \end{bmatrix}, \quad \mb{P} = \begin{bmatrix}
        0 &\\
        0 &\\
        1 &\\
        & 1
    \end{bmatrix},
\end{align*}
where it holds again that $\mb{W} = \mb{Q}$ and $\mb{V} = \mb{P}$ due to symmetry. We now list the remaining basis functions, omitting the intermediate steps for brevity,
\begin{align*}
    \mbb{Q} = \mbb{W} = \begin{bmatrix}
        1\\
        1
    \end{bmatrix}, \quad \mbb{P} = \mbb{V} = \begin{bmatrix}
        1\\
        0
    \end{bmatrix}, \quad \mbt{Q} = \begin{bmatrix}
        1\\
        0
    \end{bmatrix}, \quad \mbt{P} = \begin{bmatrix}
        0\\
        1
    \end{bmatrix}, \quad \mbt{W} = \begin{bmatrix}
        0\\
        1
    \end{bmatrix}, \quad \mbt{V} = \begin{bmatrix}
        1\\
        0
    \end{bmatrix}
\end{align*}
and finally obtain an ODE in one differential variable only, together with a purely algebraic system that recovers the remaining algebraic variables
\begin{align}
    C \ddt \varphi_3 + g_\mr{D}(\varphi_3) \varphi_3 - i_\mr{s}(t) = 0, \quad \begin{bmatrix}
        \varphi_1\\
        \varphi_2\\
        i_\mr{L}
    \end{bmatrix} = \begin{bmatrix}
        \varphi_2 + R i_\mr{s}(t)\\
        \varphi_3 + L \ddt i_\mr{s}(t)\\
        i_\mr{s}(t)
    \end{bmatrix} \label{eq:itc_do2}.
\end{align}

We note that all basis functions of the second step are also constant for this example, as was the case for the first step. As hinted at in \autoref{sec:di}, it is possible to find a purely topological decoupling based on the dissection index\cite{jansen2014}, which agrees with the intuition given by the index result from \autoref{th:ic_imna}. When considering large circuits, this topological decoupling along with its topological basis functions is to be preferred over other basis function choices, as it avoids the numerical computation of the basis functions, which becomes prohibitively expensive for large systems. We also note that a similar topological decoupling result exists for circuits including controlled sources, however only for the first step of the decoupling, as it is framed in the context of semi-explicit methods for which one only requires a DAE in semi-explicit form\cite{jansen2014}. An extension of this result to circuits of higher index is within the scope of further research.

\section{Index-aware learning}
\label{sec:ial}
\begin{figure}[b]
    \begin{center}
        \includegraphics[width=.7\textwidth]{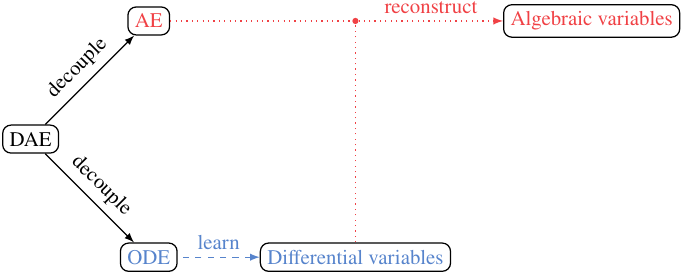}
    \end{center}
    \caption{Schematic workflow of index-aware learning.}
    \label{fig:ial_ial}
\end{figure}
In the following, we outline the use of the dissection index in the context of machine learning. The workflow is illustrated in \autoref{fig:ial_ial} and directly follows the structure of the dissection index. We give a general description in a first step, followed by examples using the two circuits from \autoref{fig:itc_do1} and \autoref{fig:itc_do2}.
\begin{enumerate}
    \item Our approach begins by performing the decoupling of a given DAE into an ODE and a purely algebraic equation (AE) following the main steps of the dissection index.
    \item Afterwards, only the differential variables of the ODE are learned. For a DAE of index one, these would be the entries of $\mbt{x}$, and for a DAE of index two, the entries of $\mbt{x}_\mb{Q}$ using the notation of \autoref{sec:di}.
    \item The remaining algebraic variables, $\mbb{x}$ for index one or $[\mbt{x}_\mb{P}, \mbb{x}_\mb{P}, \mbb{x}_\mb{Q}]^\T$ for index two, may then be reconstructed using the algebraic equations.
\end{enumerate}
We remark that the identification of the differential variables, and thus the advantages of the second point, are in principle available for any Spice based simulator via the purely topological decoupling\cite{jansen2014}, whereas the reconstruction of the algebraic variables requires an additional implementation. We summarize the important steps of this additional implementation for the index one and two cases below.

\paragraph{Index one case}
In order to recover the algebraic variables at time $t$ in the index one case, we only need to solve \eqref{eq:ioc_iob} for $\mbb{x}(t)$ using the learned $\mbt{x}(t)$.

\paragraph{Index two case}
In the index two case, we start by solving \eqref{eq:itc_xtp} for $\mbt{x}_\mb{P}(t)$ using the learned $\mbt{x}_\mb{Q}(t)$. Since we also require the derivative $\ddt \mbt{x}_\mb{P}(t)$ in \eqref{eq:itc_itiiib}, we consider a small time increment $\Delta t$ and approximate the derivative using a backward difference
\begin{align*}
    \ddt \mbt{x}_\mb{P}(t) \approx \frac{\mbt{x}_\mb{P}(t + \Delta t) - \mbt{x}_\mb{P}(t)}{\Delta t}.
\end{align*}
We note that this only reflects our implementation; in principle any finite difference (or similar) approximation is possible. Finally, we determine $\mbb{x}_\mb{P}(t)$ and $\mbb{x}_\mb{Q}(t)$ by jointly solving \eqref{eq:itc_xbp} and \eqref{eq:itc_itiiib} using the learned $\mbt{x}_\mb{Q}(t)$, $\mbt{x}_\mb{P}(t)$ and the approximation of $\ddt \mbt{x}_\mb{P}(t)$.

\paragraph{Example circuits}
In terms of the example circuits from \autoref{sec:di}, the workflow amounts to the following: for the first example we consider $\varphi_3$ and $i_\mr{L}$ as the differential variables that need to be learned, and for the second example only $\varphi_3$ is left. All the remaining algebraic variables may be recovered using \eqref{eq:itc_do1} or \eqref{eq:itc_do2} respectively. Thus the learning effort is already reduced quite significantly in these two examples; from five to two variables in the first and from four down to one variable in the second. Another key benefit comes from the fact that the reconstructed algebraic variables exactly fulfill the inherent constraints of the DAE. This means that even though the learned solution variables (think of $\varphi_3$ for example) are only approximations, the reconstructions (e.g.~$\varphi_1$ or $\varphi_2$) will still be consistent. This may be of great importance for systems where the physical interpretability of the solution depends on it satisfying the constraints.

There is yet another, maybe less expected, benefit that might occur. Looking at the decoupled system from \eqref{eq:itc_do2} we find that the resistance parameter $R$ and the inductance parameter $L$ only appear in the algebraic equation. In terms of our original goal of speeding up design optimization or uncertainty quantification, where a lot of solutions for varying parameter values are required, this means the following: instead of having to solve the full system for a given combination of $R$ and $L$, we can instead simply solve the algebraic equation to obtain the full solution. While the algebraic equation might be more complicated than the simple linear relations of \eqref{eq:itc_do1} and \eqref{eq:itc_do2} in general, solving it is almost certainly much faster than having to integrate the entire system in time. This becomes an even bigger advantage when knowledge about the solution is only required for specific points in time, since the algebraic equation may be solved pointwise. As of now, we have no easy way to automatically determine which parameters appear in the ODE. But when combined with a sequential learning strategy, such as the one outlined in \autoref{sec:ne}, there may still be computational savings due to the learning method requiring less samples for the parameters not appearing in the ODE.

Lastly, we emphasize that the approach is independent of the particular machine learning method that is used for learning the differential variables. Thus methods developed especially for ODEs may be employed and exchanged depending on the problem at hand.

\section{Numerical examples}
\label{sec:ne}
Before we present numerical results, we want to provide some background on our machine learning method of choice, Gaussian processes (GPs), and the particular learning strategy we employ.

\paragraph{Gaussian processes}
The following brief introduction is based on the textbook of Rasmussen and Williams\cite{rasmussen2006}, and we refer to the book itself for more details. We consider the problem of learning one component $x(t)$ of the DAE solution $\mb{x}(t)$, compare \eqref{eq:di_dae}, based on observations
\begin{align*}
    O = \big\{ (t_i, x_i): 1 \leq i \leq N \big\}.
\end{align*}
We will focus on the one-dimensional case for clarity of exposition, however we remark that the ideas extend to the case where the solution component $x(t, \mb{p})$ also depends on the parameters $\mb{p}$ and thus more than one variable, compare also the textbook\cite{rasmussen2006}. A GP suited for this problem is defined by a (prior) mean function $m: \mathbb{R} \to \mathbb{R}$ and covariance function $k: \mathbb{R} \times \mathbb{R} \to \mathbb{R}$. The learning problem is then tackled using Bayesian inference, such that one aims to obtain the posterior distribution $\hat{x}(t)$, given the observations $O$ and a point $t$, where $x$ is to be predicted. A particular feature of GPs is that this posterior process, under suitable assumptions, turns out to be another GP with posterior mean and covariance\cite{basak2021}
\begin{subequations}
    \label{eq:ne_gp}
    \begin{align}
        \hat{m}(t) &= m(t) + \mb{w}(t)^\T (\mb{x} - \mb{m}) \label{eq:ne_gpa}\\
        \hat{k}(t) &= k(t, t) - \mb{w}(t)^\T \mb{k}(t) \label{eq:ne_gpb},
    \end{align}
\end{subequations}
where $\mb{m} \coloneqq [m(t_1), \dotsc, m(t_N)]^\T$ denotes the prior mean function evaluated at the observations, $\mb{k}(t) \coloneqq [k(t, t_1), \dotsc, k(t, t_N)]^\T$ similarly denotes the pairwise evaluation of the covariance function using the prediction point $t$ and the observations, and $\mb{x} \coloneqq [x_1, \dotsc, x_N]^\T$ are the observed function values. The weights $\mb{w}(t)$ are given by the solution of
\begin{align*}
    \big( \mb{K} + \sigma^2 \mb{I} \big) \mb{w}(t) = \mb{k}(t),
\end{align*}
where $[\mb{K}]_{i,j} \coloneqq k(t_i, t_j)$ and $\sigma^2$ models i.i.d.~additive Gaussian noise on the observations. For a discussion on modeling choices where the assumptions leading to \eqref{eq:ne_gp} are not fulfilled, see the review article by Swiler et al.\cite{swiler2020}.

The key model component influencing the learning process is the covariance (or kernel) function $k$, since it determines the approximation properties of the GP. It encodes prior knowledge about the function $x(t)$ that is to be learned, such as its differentiability or characteristic length scales. We opt for a radial basis function kernel, given by
\begin{align*}
    k(t_i, t_j, \sigma_k, \ell) = \sigma_k^2 \exp \biggl(- \Big( \frac{t_i - t_j}{\ell} \Big)^2 \biggr),
\end{align*}
where $\sigma_k$ and the length scale $\ell$ are hyperparameters, which allow for better approximation capabilities of the GP. The kernel is selected to match the differentiability of the solution.

In practice, the mean function $m$ is often taken to be zero as the data is assumed standardized, and the hyperparameters are then determined by minimizing the negative log likelihood\cite{basak2021}
\begin{align*}
    -\log \bigl( p(\mb{x} | \sigma, \sigma_k, \ell) \bigr) = \frac{1}{2} \big( \mb{x}^\T \mb{K}(\sigma, \sigma_k, \ell)^{-1} \mb{x} + \log \bigl( \det \mb{K}(\sigma, \sigma_k, \ell) \bigr) + N \log(2\pi) \big),
\end{align*}
where $[\mb{K}(\sigma, \sigma_k, \ell)]_{i,j} \coloneqq k(t_i, t_j, \sigma_k, \ell) + \sigma^2$.

\paragraph{Learning strategy}
The learning strategy aims to exploit one of the key features of GPs: they provide both a mean prediction $\hat{m}$ and an associated variance estimate $\hat{k}$, as detailed in \eqref{eq:ne_gp}. We use these properties in conjunction with a sequential sample selection strategy, that starts out with a small number of training data and adds further samples based on the variance estimate of the GP. This idea is not new, see again the textbook of Rasmussen and Williams\cite{rasmussen2006} for more references and details, however we still want to outline our particular approach to make the results better interpretable and reproducible.

Our implementation proceeds as follows:
\begin{enumerate}
    \item We select a grid of time points $T = \{ t_i: 1 \leq i \leq N_T \}$ and parameter values $P = \{ \mb{p}_i: 1 \leq i \leq N_P \}$, where each $\mb{p}_i$ represents a specific combination of parameter values, for which we want to learn the solution of the DAE using a GP.
    \item We select a subset $D \subset T \times P$ and use the corresponding solutions for the initial training of a GP.
    \item As a termination criterion, we compute the posterior mean $\hat{m}(t, \mb{p})$ using \eqref{eq:ne_gpa} in all grid points $(t, \mb{p}) \in T \times P$ and check whether the relative prediction error
    \begin{align*}
        e \coloneqq \frac{\| \mbh{m} - \mb{x} \|_2}{ \| \mb{x} \|_2}
    \end{align*}
    is below a desired tolerance, and if not, continue with 4. Here, $\mbh{m} \coloneqq \big[ \hat{m}(t_1, \mb{p}_1), \dotsc, \hat{m}(t_{N_T}, \mb{p}_{N_P}) \big]$ collects the mean predictions for all $(t, \mb{p}) \in T \times P$ and $\mb{x}$ contains the corresponding simulated solution values.
    \item We compute the variance prediction $\hat{k}(t, \mb{p})$ using \eqref{eq:ne_gpb} for all $(t, \mb{p}) \in (T \times P) \setminus D$ and add a point of maximum variance to the training data set $D$.
    \item Finally, we retrain the GP and continue with 3.
\end{enumerate}

We note that several improvements may be made to this strategy, such as properly maximizing the variance estimate in 4, instead of sampling on a discrete grid. The implementation is based on the STK toolbox\cite{stk}.

\paragraph{First example circuit}
\begin{figure}[t]
    \begin{center}
        \includegraphics[width=.4\textwidth]{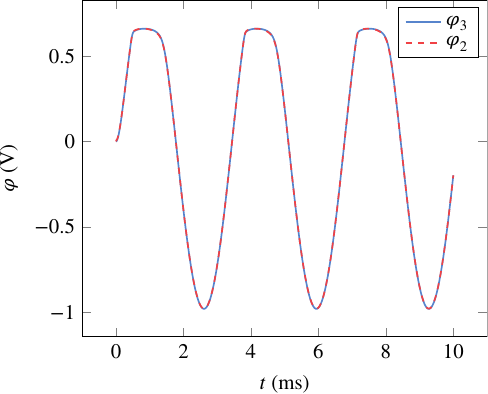} \hspace{1.5cm} \includegraphics[width=.4\textwidth]{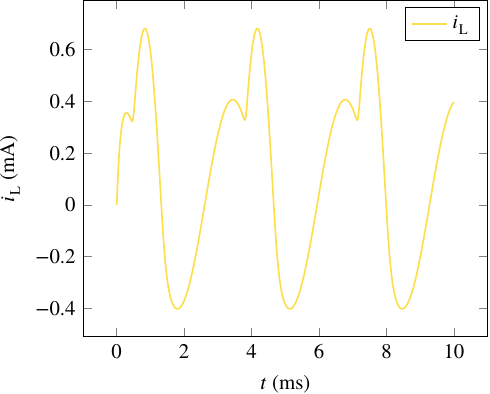}
    \end{center}
    \caption{Solution of the first example circuit for $L = \SI{1.7}{\milli\henry}$ and $C = \SI{220}{\nano\farad}$.}
    \label{fig:ne_do1_solution}
\end{figure}
We again consider the example of \autoref{fig:itc_do1}, where we choose $v_\mr{s}(t) = \sin(600 \pi t)\, \si{\volt}$, $R = \SI{500}{\ohm}$ and the diode is modeled by $g_\mr{D}(\varphi_3) = 10^{-14} \big( e^{(\varphi_3 / \SI{26}{\mV})} - 1 \big)\, \si{\siemens}$. The remaining parameter values are chosen according to the sequential sample selection strategy with $\SI{1}{\milli\henry} \leq L \leq \SI{3}{\milli\henry}$ and $\SI{100}{\nano\farad} \leq C \leq \SI{300}{\nano\farad}$, i.e.~$\mb{p} = [L, C]^\T$ in the context of \autoref{sec:int}. The starting points for the strategy are given by all combinations of the boundary values for $L$ and $C$ together with $t = \SI{0}{\ms}$ and $t = \SI{10}{\ms}$ once for each combination. We note that the choice to consider $L$ and $C$, and not for example $R$, as parameters here is arbitrary and only serves to illustrate the approach. One could still follow the same approach when considering e.g.~the initial conditions, or the diode model, as being parameterized. Recalling \eqref{eq:itc_do1}, we see that $\varphi_3$ and $i_\mr{L}$ have to be learned as the differential variables. In addition to these two, we also consider the algebraic variable $\varphi_2$ and we then compare the accuracy of learning $\varphi_2$ directly to recovering it from the two differential variables. \autoref{fig:ne_do1_solution} shows the solution for $L = \SI{1.7}{\milli\henry}$ and $C = \SI{220}{\nano\farad}$, a combination which is not part of the training data.
\begin{figure}[b]
    \begin{center}
        \includegraphics[width=.4\textwidth]{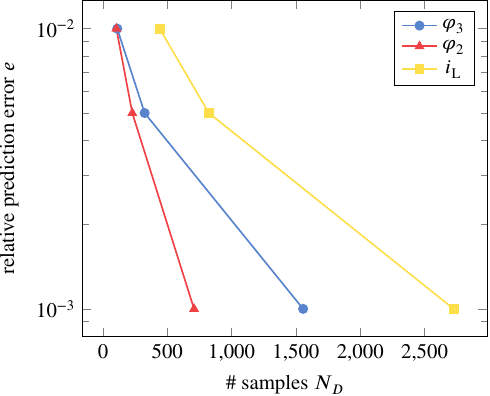}
    \end{center}
    \caption{Convergence of the relative prediction error $e$, for the variables shown in \autoref{fig:ne_do1_solution}. The final accuracies correspond to approximately 180 different combinations of $L$ and $C$ for $\varphi_3$, 39 combinations for $\varphi_2$ and 37 for $i_\mr{L}$.}
    \label{fig:ne_do1_convergence}
\end{figure}

In \autoref{fig:ne_do1_convergence}, we see the convergence of the relative prediction errors with respect to the number of samples used by the sample selection strategy. We observe that, depending on the qualitative complexity of the dynamics, the solution variables take different numbers of samples to reach the same prediction accuracy. It should be noted however, that the total number of samples $N_D$ also includes the number of time points that are sampled, and not only the number of distinct parameter combinations (simulations). The latter are listed in the caption of \autoref{fig:ne_do1_convergence} and turn out to be considerably smaller. One may also note that we only execute the learning strategy up to a relatively large tolerance of $10^{-3}$. This is due to the fact that both the optimization of the hyperparameters, as well as the computation of the posterior mean and variance, scale badly for conventional GPs, leading to large computation times. Remedies for this exist, see e.g.~the book by Rasmussen and Williams\cite{rasmussen2006}, however this issue lies outside the scope of this article as our approach may be combined with any machine learning method of choice.
\begin{figure}[t]
    \begin{center}
        \includegraphics[width=.4\textwidth]{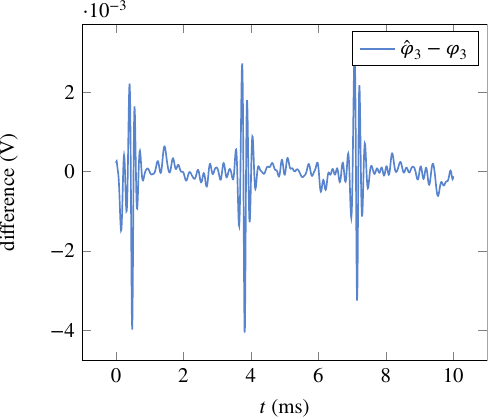} \hspace{1.5cm} \includegraphics[width=.4\textwidth]{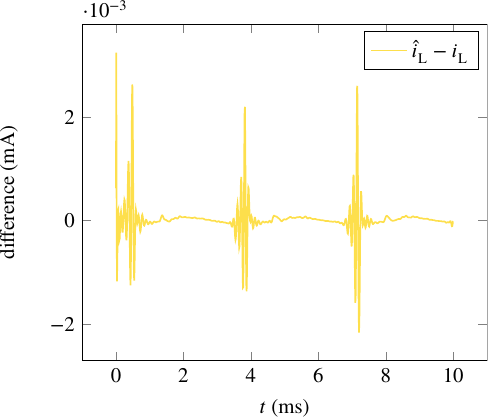}
    \end{center}
    \caption{Differences between the mean predictions ($\hat{\varphi}_3$ and $\hat{i}_\mr{L}$) and the corresponding simulations for the differential variables when considering $L = \SI{1.7}{\milli\henry}$ and $C = \SI{220}{\nano\farad}$. The predictions correspond to the smallest relative errors from \autoref{fig:ne_do1_convergence}.}
    \label{fig:ne_do1_error_d}
\end{figure}

\begin{figure}[b]
    \begin{center}
        \includegraphics[width=.4\textwidth]{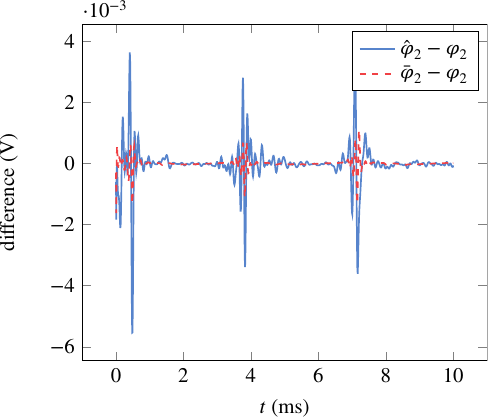} \hspace{1.5cm} \includegraphics[width=.4\textwidth]{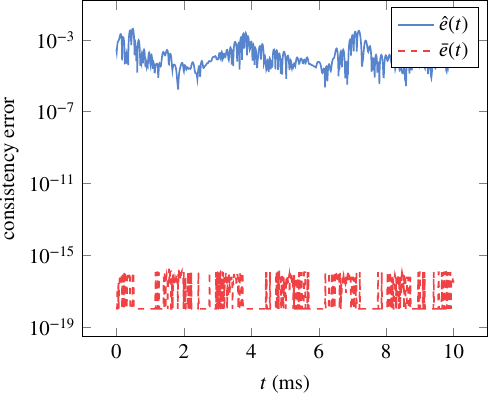}
    \end{center}
    \caption{Differences between the mean prediction $\hat{\varphi}_2$, reconstruction $\bar{\varphi}_2$ and the corresponding simulation of the algebraic variable $\varphi_2$ (left) and the respective consistency errors (right). The predictions correspond to $L = \SI{1.7}{\milli\henry}$, $C = \SI{220}{\nano\farad}$ and the smallest relative error from \autoref{fig:ne_do1_convergence}.}
    \label{fig:ne_do1_error_a_consistency}
\end{figure}
The differences between the mean predictions and simulations for the differential variables, when using the predictions belonging to the smallest relative errors from \autoref{fig:ne_do1_convergence}, are shown in \autoref{fig:ne_do1_error_d}. The results again correspond to $L = \SI{1.7}{\milli\henry}$ and $C = \SI{220}{\nano\farad}$, and we observe that the differences are in line with the relative prediction errors of \autoref{fig:ne_do1_convergence}. The differences between the mean prediction $\hat{\varphi}_2$, reconstruction $\bar{\varphi}_2$ and the simulation of the algebraic variable $\varphi_2$ are highlighted on the left of \autoref{fig:ne_do1_error_a_consistency}. The results again correspond to the same parameter values $L = \SI{1.7}{\milli\henry}$ and $C = \SI{220}{\nano\farad}$. We observe that there is not much difference between the mean prediction and reconstruction, however the reconstruction does appear to have a slight advantage in terms of accuracy. Here, one should note that although the reconstruction is exact up to the accuracy of solving the algebraic equation in \eqref{eq:itc_do1}, it still contains the error from learning the differential variables, hence the overall difference between $\bar{\varphi}_2$ and $\varphi_2$. To better quantify the difference between the learned and reconstructed solutions, we introduce consistency errors $\hat{e}(t)$ and $\bar{e}(t)$ based on \eqref{eq:itc_do1}
\begin{align*}
    \hat{e}(t) \coloneqq \begin{Vmatrix}
        \hat{\varphi}_1 - v_\mr{s}(t)\\
        \hat{\varphi}_2 + R \hat{i}_\mr{L} - v_\mr{s}(t)\\
        \hat{i}_\mr{V} + \hat{i}_\mr{L}
    \end{Vmatrix}_2, \quad \bar{e}(t) \coloneqq \begin{Vmatrix}
        \bar{\varphi}_1 - v_\mr{s}(t)\\
        \bar{\varphi}_2 + R \hat{i}_\mr{L} - v_\mr{s}(t)\\
        \bar{i}_\mr{V} + \hat{i}_\mr{L}
    \end{Vmatrix}_2,
\end{align*}
where $\hat{\cdot}$ indicates learned variables and $\bar{\cdot}$ refers to reconstructed variables. For a consistent solution obeying all algebraic constraints, both errors are identically zero. The right of \autoref{fig:ne_do1_error_a_consistency} shows these consistency errors when using the same predictions and reconstructions as on the left. We observe a very clear improvement of the reconstructions (consistency error on the order of machine precision) over the directly learned predictions (consistency error as large as $10^{-3}$). While the central benefit is this adherence to the constraints, only having to learn two of the five solution variables also represents a significant reduction of the learning effort in this case.

\begin{figure}[b]
    \begin{center}
        \includegraphics[width=.4\textwidth]{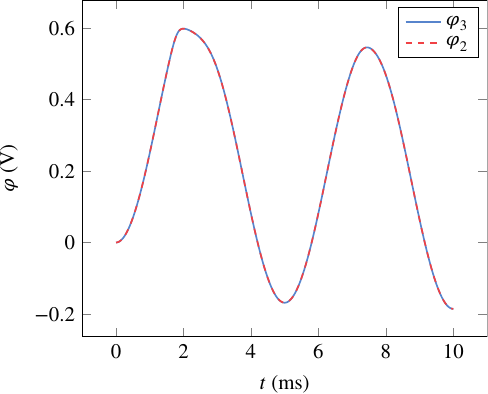} \hspace{1.5cm} \includegraphics[width=.4\textwidth]{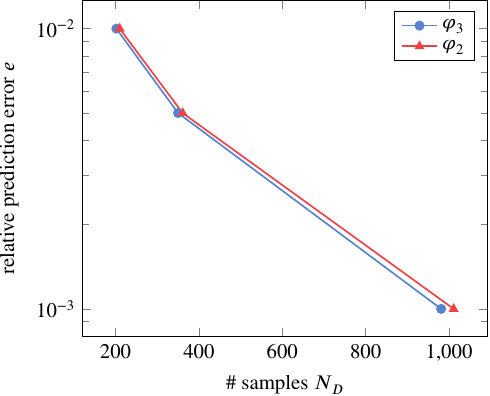}
    \end{center}
    \caption{Solution of the second example circuit for $L = \SI{1.7}{\milli\henry}$ and $C = \SI{220}{\nano\farad}$ (left) and convergence of the relative prediction error for the same variables (right). The final accuracies correspond to 31 distinct combinations of $L$ and $C$ for $\varphi_3$ and 49 for $\varphi_2$.}
    \label{fig:ne_do2_solution_convergence}
\end{figure}

\paragraph{Second example circuit}
For the second example from \autoref{fig:itc_do2}, we select $i_\mr{s}(t) = 10^{-4} \sin(400 \pi t)\, \si{\volt}$, while all other parameters and the learning strategy remain the same. Recalling \eqref{eq:itc_do2}, we find that $\varphi_3$ is the only differential variable that needs to be learned to reconstruct the remaining algebraic variables. We again focus on $\varphi_2$ as an algebraic variable to obtain a comparative example. Considering similar numerical studies as for the first example, the left panel of \autoref{fig:ne_do2_solution_convergence} shows the solutions of $\varphi_3$ and $\varphi_2$ for $L = \SI{1.7}{\milli\henry}$ and $C = \SI{220}{\nano\farad}$, which are again not part of the training data. The right plot shows the convergence of the relative prediction errors with respect to the total number of samples $N_D$. We again note that the number of distinct parameter combinations is significantly smaller, as listed in the caption. We also see that the relative errors behave similar for both variables in this case, which is to be expected given the similar outlook of their dynamics in the left plot.

At this point we also return to the discussion from \autoref{sec:ial} about parameters only appearing in the algebraic equation. During the learning process, the sample selection strategy requested 21 unique values for $C$, all of which required full simulations according to \eqref{eq:itc_do2}. The 7 unique values for $L$, that were requested for learning $\varphi_2$, did not require full simulations however, but rather could be reconstructed from \eqref{eq:itc_do2}. This results from the fact that the ODE in \eqref{eq:itc_do2} does not depend on $L$, such that the solution of $\varphi_3$ also does not depend on that parameter. The same goes for the algebraic variable $i_\mr{L}$, thus $\varphi_2$ and $\varphi_1$ can be reconstructed only based on the knowledge of $\varphi_3$. We again emphasize that time is also included as a parameter within the sample selection strategy, such that the reconstructions of the algebraic variables only need to be evaluated at the particular points in time that are requested by the strategy, rather than at all time points of the solution.
\begin{figure}[t]
    \begin{center}
        \includegraphics[width=.4\textwidth]{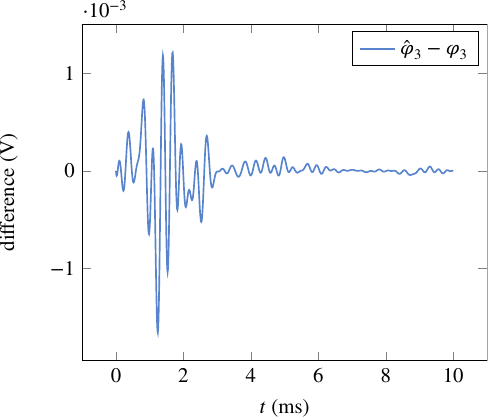} \hspace{1.5cm} \includegraphics[width=.4\textwidth]{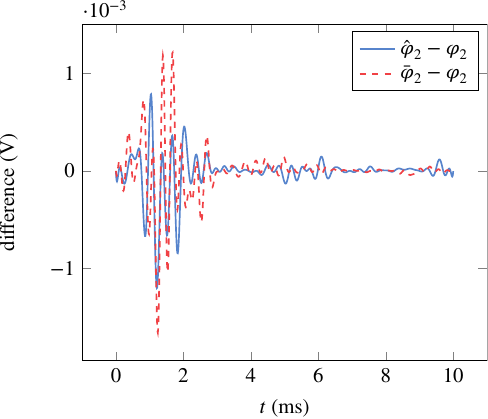}
    \end{center}
    \caption{Differences between the mean prediction and simulation for $\varphi_3$ (left) and between the mean prediction, reconstruction and simulation for $\varphi_2$ (right). The predictions are again made for $L = \SI{1.7}{\milli\henry}$ and $C = \SI{220}{\nano\farad}$ and correspond to the smallest relative errors from \autoref{fig:ne_do2_solution_convergence}.}
    \label{fig:ne_do2_error}
\end{figure}

\begin{figure}[b]
    \begin{center}
        \includegraphics[width=.4\textwidth]{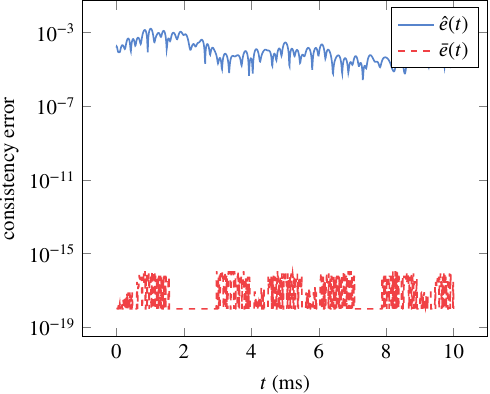}
    \end{center}
    \caption{Consistency errors $\hat{e}(t)$ and $\bar{e}(t)$ corresponding to the results from \autoref{fig:ne_do2_error}.}
    \label{fig:ne_do2_consistency}
\end{figure}

The difference between the mean prediction and simulation of $\varphi_3$, for $L = \SI{1.7}{\milli\henry}$ and $C = \SI{220}{\nano\farad}$, is shown on the left of \autoref{fig:ne_do2_error}, while the right shows the differences between the mean prediction, reconstruction and simulation of $\varphi_2$. The predictions again correspond to the smallest relative errors from \autoref{fig:ne_do2_solution_convergence}, and the differences are of the same order. In this case, the reconstruction performs similar or even slightly worse compared to the mean prediction when only looking at the difference. However it still constitutes a reduction in learning effort, from four solution variables down to only one, and most importantly the reconstruction adheres to the algebraic constraints of the DAE. To further illustrate this point, we again take a look at the consistency errors, now redefined based on \eqref{eq:itc_do2}
\begin{align*}
    \hat{e}(t) \coloneqq \begin{Vmatrix}
        \hat{\varphi}_1 - \hat{\varphi}_2 + R i_\mr{s}(t)\\
        \hat{\varphi}_2 - \hat{\varphi}_3 + L \ddt i_\mr{s}(t)\\
        \hat{i}_\mr{L} - i_\mr{s}(t)
    \end{Vmatrix}_2, \quad \bar{e}(t) \coloneqq \begin{Vmatrix}
        \bar{\varphi}_1 - \bar{\varphi}_2 + R i_\mr{s}(t)\\
        \bar{\varphi}_2 - \hat{\varphi}_3 + L \ddt i_\mr{s}(t)\\
        \bar{i}_\mr{L} - i_\mr{s}(t)
    \end{Vmatrix}_2.
\end{align*}
In our example, the algebraic equation from \eqref{eq:itc_do2} gives an explicit description of the algebraic variables once more, such that the reconstruction is accurate up to machine precision, compare $\bar{e}(t)$ in \autoref{fig:ne_do2_consistency}. When learning all solution variables individually however, we observe a much larger maximum value of around $10^{-3}$ for the consistency error $\hat{e}(t)$ across all the predicted points in time. In general, the algebraic equation may be nonlinear such that the reconstruction still leads to a consistency error $\bar{e}(t)$ greater than machine precision, depending on the accuracy of the nonlinear system solver.

\paragraph{Rectifier circuit}
\begin{figure}[t]
    \begin{center}
        \includegraphics[width=0.5\textwidth]{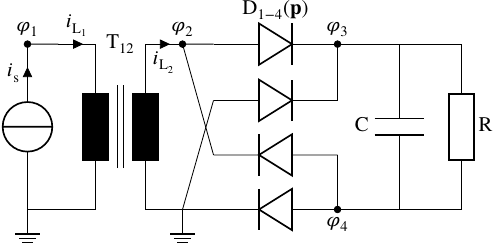}
    \end{center}
    \caption{Full wave rectifier circuit. The resistances of the diodes depend on the parameters $\mb{p} = [T_1, T_2]^\T$.}
    \label{fig:ne_fwr2}
\end{figure}

\begin{figure}[b]
    \begin{center}
        \includegraphics[width=.4\textwidth]{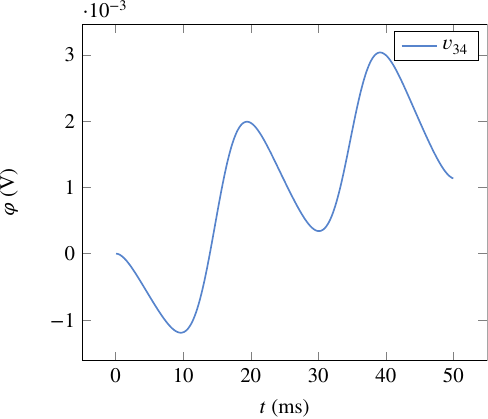} \hspace{1.5cm} \includegraphics[width=.4\textwidth]{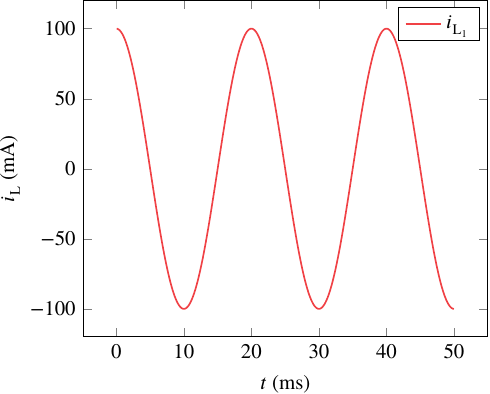}
    \end{center}
    \caption{Solutions of the differential variables of the full wave rectifier circuit for $T_1 = 65\, \si{\celsius}$ and $T_2 = 85\, \si{\celsius}$.}
    \label{fig:ne_fwr2_solution_d}
\end{figure}
As a third and larger example we consider the rectifier circuit from \autoref{fig:ne_fwr2}. Aside from requiring the full index two implementation from \autoref{sec:ial}, this example also showcases another potential application in the context of electrothermal simulations. Electrothermal simulations are used to investigate the thermal behavior of a circuit, by coupling the power that is dissipated in the circuit to a set of equations describing the temperature distribution in the circuit, and by using the temperatures of some components as parameters for certain device functions. In our case, we again model the diodes as nonlinear resistors, however this time the model also includes a temperature dependence\cite{tietze2008}
\begin{align*}
    g_\mr{D}(v_\mr{D}, T_\mr{D}) = I_\mr{s}(T_\mr{D}) \frac{qk}{T_\mr{D}} e^{(qk/T_\mr{D}) v_\mr{D}},
\end{align*}
where $v_\mr{D}$ is the voltage across the diode in forward direction, $T_\mr{D}$ is the temperature of the diode, $q$ is the charge of an electron, $k$ the Boltzmann constant and $I_\mr{s}(T_\mr{D})$ the temperature dependent reverse saturation current\cite{tietze2008}. The current source is given by $i_\mr{s}(t) = 0.1 \cos(100 \pi t)\, \si{\ampere}$ and the capacitor and resistor are linear with $C = 1\, \si{\milli\farad}$, $R = 50\, \si{\ohm}$. The circuit also contains a transformer $\mr{T}_{12}$ modeled as a nonlinear inductive multiport with
\begin{align*}
    \mb{L}(\i{L}) = \begin{bmatrix}
        L(i_{\mr{L}_1}) & \sqrt{0.09 L(i_{\mr{L}_1}) L(i_{\mr{L}_2})}\\
        \sqrt{0.09 L(i_{\mr{L}_1}) L(i_{\mr{L}_2})} & 0.1 L(i_{\mr{L}_2})
    \end{bmatrix},
\end{align*}
where $L(i_\mr{L})$ is a fourth order polynomial in $i_\mr{L}$\cite{lullo2017}. We also note that rectifiers play an important role in many power electronics applications; they are key when converting high AC to lower DC voltages. In the example we still use a current source, since this leads to an index two circuit in this case, compare \autoref{th:ic_imna}, which requires the full index two approach from \autoref{sec:ial}.
\begin{figure}[t]
    \begin{center}
        \includegraphics[width=.4\textwidth]{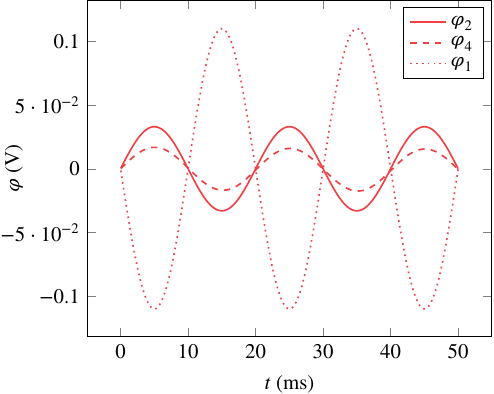} \hspace{1.5cm} \includegraphics[width=.4\textwidth]{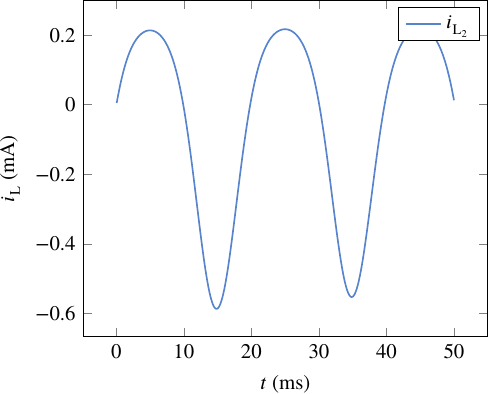}
    \end{center}
    \caption{Solutions of the algebraic variables of the full wave rectifier circuit for $T_1 = 65\, \si{\celsius}$ and $T_2 = 85\, \si{\celsius}$.}
    \label{fig:ne_fwr2_solution_a}
\end{figure}

\begin{figure}[b]
    \begin{center}
        \includegraphics[width=.4\textwidth]{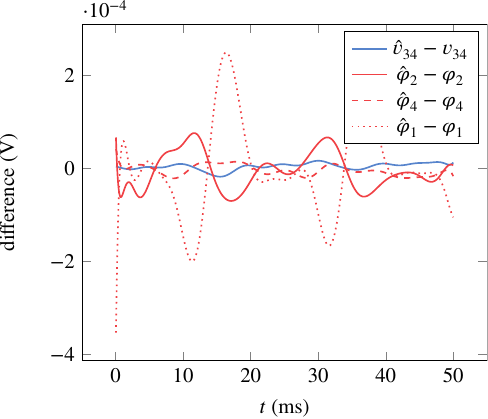} \hspace{1.5cm} \includegraphics[width=.4\textwidth]{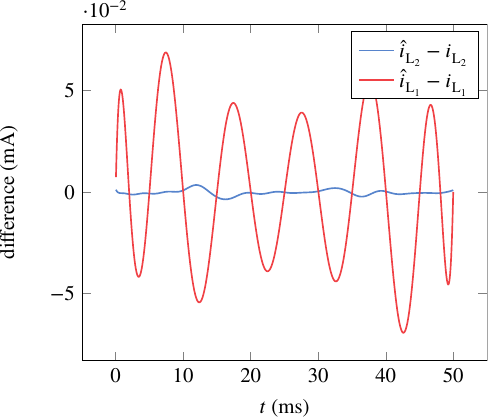}
    \end{center}
    \caption{Differences between the mean predictions ($\hat{\cdot}$) and simulations of all differential and algebraic variables for $T_1 = 65\, \si{\celsius}$ and $T_2 = 85\, \si{\celsius}$.}
    \label{fig:ne_fwr2_error}
\end{figure}
When decoupling the DAE arising from the circuit of \autoref{fig:ne_fwr2} using the dissection index, one finds
\begin{align*}
    \mbt{x}_\mb{Q} = \begin{bmatrix}
        v_{34}\\
        i_{\mr{L}_2}
    \end{bmatrix}, \quad \mbt{x}_\mb{P} = i_{\mr{L}_1}, \quad \mbb{x}_\mb{P} = \begin{bmatrix}
        \varphi_2\\
        \varphi_4
    \end{bmatrix}, \quad \mbb{x}_\mb{Q} = \varphi_1
\end{align*}
for the differential and algebraic variables using the notation from \autoref{sec:di}. We first observe that there are only two differential variables $\mbt{x}_\mb{Q}$, again leading to a significant reduction in the learning effort. We also see that the first differential variable $v_{34} \coloneqq \varphi_3 - \varphi_4$ is a linear combination of the original variables of the DAE. This is a general phenomenon and does not interfere with our approach, as the original variables may always be reconstructed from the sets of differential and algebraic variables by reversing the splitting using \eqref{eq:ioc_x}, \eqref{eq:itc_xb} and \eqref{eq:itc_xt} from \autoref{sec:di}
\begin{align*}
    \mb{x} = \mb{P} \mbt{x} + \mb{Q} \mbb{x} = \mb{P} \big( \mbt{P} \mbt{x}_\mb{P} + \mbt{Q} \mbt{x}_\mb{Q} \big) + \mb{Q} \big( \mbb{P} \mbb{x}_\mb{P} + \mbb{Q} \mbb{x}_\mb{Q} \big).
\end{align*}
To keep the overall learning effort manageable, we consider the outer diodes $\mr{D}_1$ and $\mr{D}_4$ to depend on the same temperature $T_1$ and the inner diodes $\mr{D}_2$ and $\mr{D}_3$ to depend on the same temperature $T_2$, such that the vector of parameters is now given by $\mb{p} = [T_1, T_2]^\T$. We emphasize that there is no need to restrict the vector of parameters to two entries, this choice is only made to reduce the learning effort. For the training data we consider temperatures $\SI{20}{\celsius} \leq T_1, T_2 \leq \SI{90}{\celsius}$, and the starting values for the sample selection strategy are again obtained by considering all combinations of the boundary values together with $t = \SI{0}{\ms}$ and $t = \SI{50}{\ms}$. We then execute the learning strategy up to a tolerance of $5 \cdot 10^{-3}$ and choose the prediction point $T_1 = \SI{65}{\celsius}$, $T_2 = \SI{85}{\celsius}$ that is not included in the training data.

\begin{figure}
    \begin{center}
        \includegraphics[width=.4\textwidth]{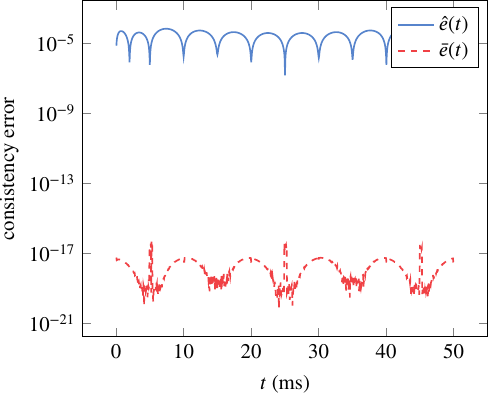}
    \end{center}
    \caption{Consistency errors $\hat{e}(t)$ and $\bar{e}(t)$ corresponding to the results from \autoref{fig:ne_fwr2_error}}
    \label{fig:ne_fwr2_consistency}
\end{figure}

The solutions of the differential and algebraic variables, for $T_1 = 65\, \si{\celsius}$ and $T_2 = 85\, \si{\celsius}$, are shown in \autoref{fig:ne_fwr2_solution_d} and \autoref{fig:ne_fwr2_solution_a}, while \autoref{fig:ne_fwr2_error} shows the differences between the mean predictions and simulations for the same temperature values. The comparatively larger differences for $i_{\mr{L}_1}$ in \autoref{fig:ne_fwr2_error} stem from the prediction error being defined as a relative quantity and the absolute values of $i_{\mr{L}_1}$ being much larger than the others, compare \autoref{fig:ne_fwr2_solution_d}. Defining a function $\mb{g}(\mbt{x}_\mb{Q}, \mbt{x}_\mb{P}, \mbb{x}_\mb{P}, \mbb{x}_\mb{Q}, t, \mb{p})$ by stacking the nonlinear systems \eqref{eq:itc_xtp}, \eqref{eq:itc_xbp} and \eqref{eq:itc_itiiib} required to recover the algebraic variables, compare \autoref{sec:ial}, we can again define consistency errors
\begin{align*}
    \hat{e}(t) \coloneqq \big\| \mb{g}(\hat{\mbt{x}}_\mb{Q}, \hat{\mbt{x}}_\mb{P}, \hat{\mbb{x}}_\mb{P}, \hat{\mbb{x}}_\mb{Q}, t, \mb{p}) \big\|_2, \quad \bar{e}(t) \coloneqq \big\| \mb{g}(\hat{\mbt{x}}_\mb{Q}, \bar{\mbt{x}}_\mb{P}, \bar{\mbb{x}}_\mb{P}, \bar{\mbb{x}}_\mb{Q}, t, \mb{p}) \big\|_2,
\end{align*}
where the additional $\hat{\cdot}$ and $\bar{\cdot}$ again refer to learned or reconstructed variables respectively. For $T_1 = 65\, \si{\celsius}$ and $T_2 = 85\, \si{\celsius}$ the consistency errors can be seen in \autoref{fig:ne_fwr2_consistency}. We observe that the reconstructed solution again outperforms the directly learned solution by several orders of magnitude.

\section{Conclusions and future research}
\label{sec:cfr}
This article introduced a new approach for learning the time and parameter dependent solutions of electrical circuits. The approach assumes the circuit to be modeled using MNA, and then exploits the structure of MNA to improve the learning process by splitting it into a learning and a reconstruction step, compare \autoref{fig:ial_ial}. It achieves this by decoupling the underlying DAE into an ODE and a purely algebraic equation. Benefits of the approach include a reduction in the number of variables that are to be learned during the learning step, and the exact adherence of the learned solution to the inherent constraints of the circuit model after the reconstruction step. Numerical examples illustrated both benefits. The examples also showed that the exact recovery of the constraints might improve the accuracy of the learned solutions. Furthermore, additional computational savings may be possible as some of the parameters of interest only appear in the reconstruction step, which avoids the need for training data with varying values of these parameters entirely. We emphasize that the approach is independent of the machine learning method used during the learning step, such that the learning method may be chosen according to the problem at hand.

Multiple extensions are possible within this index-aware learning framework. A natural first step could be to make use of the topological decoupling that was hinted at in \autoref{sec:di}. This would pave the way for the inclusion of controlled sources within the workflow, and thus the industrial use of the approach, with the idea being the extension of the topological decoupling to also allow for controlled sources. Focusing in on the idea of adhering to physically meaningful constraints, one could work on extending the approach to a charge conserving variant of MNA, to potentially guarantee charge conservation even for the learned solutions. Yet another direction may be the application of index-aware learning to DAEs arising elsewhere, e.g.~in modified loop analysis. (The dissection index applies to a more general class of DAEs, as \autoref{sec:di} showed.) Finally, separate but related work may focus on improving the learning step by developing new methods that are especially suited for learning ODEs.

\section*{Acknowledgments}
This work is supported by the Graduate School CE within the Centre for Computational Engineering at Technische Universit{\"a}t Darmstadt and the ECSEL Joint Undertaking (JU) under grant agreement No.~101007319. The JU receives support from the European Union's Horizon 2020 research and innovation programme and the Netherlands, Hungary, France, Poland, Austria, Germany, Italy and Switzerland. Note that this work only reflects the authors' views and that the JU is not responsible for any use that may be made of the information it contains.



\subsection*{Conflict of interest}
The authors declare no potential conflict of interests.


\appendix
\section{Dissection index and MNA}
\label{sec:dim}
\begin{remark}
    \label{re:dim_bf}
    Recalling \eqref{eq:md_mnaii}, we observe
    \begin{align*}
        \mb{M}(\mb{x}) = \begin{bmatrix}
            \A{C}^{\phantom{\T}} \mb{C} \big( \AT{C} \vphi \big) \AT{C} & &\\
            & \mb{L}(\i{L}) &\\
            & & \mb{0}
        \end{bmatrix}
    \end{align*}
    in the case of MNA. We now determine the basis functions $\mb{Q}$ and $\mb{P}$. Taking into account the standard assumption, compare \autoref{th:ic_imna}, that $\mb{L}(\i{L})$ and $\mb{C} \big( \AT{C} \vphi \big)$ are positive definite, we find\cite{jansen2014}
    \begin{align*}
        \mb{Q} = \begin{bmatrix}
            \mb{Q}_\mr{C} &\\
            & \mb{0}\\
            & \mb{I}
        \end{bmatrix}, \quad \mb{P} = \begin{bmatrix}
            \mb{P}_\mr{C} &\\
            & \mb{I}\\
            & \mb{0}
        \end{bmatrix},
    \end{align*}
    where $\mr{im\, } \mb{Q}_\mr{C} = \ker \AT{C}$, $\mb{0}$ and $\mb{I}$ are zero and identity matrices of appropriate dimensions and the columns of $\mb{Q}_\mr{C}$ and $\mb{P}_\mr{C}$ together form a basis of $\mathbb{R}^{n_{\vphi}}$ with $n_{\vphi}$ the dimension of $\vphi$. The description of $\mb{Q}_\mr{C}$ results from the fact that (with $\mb{y} \coloneqq \AT{C} \mb{x}$)
    \begin{align*}
        \mb{x}^\T \A{C}^{\phantom{\T}} \mb{C} \big( \AT{C} \vphi \big) \AT{C} \mb{x} = \mb{y}^\T \mb{C} \big( \AT{C} \vphi \big) \mb{y} > 0, \quad \forall \mb{y} \in \mathbb{R}^{n_{\vphi}} \setminus \{ \mb{0} \},
    \end{align*}
    since $\mb{C} \big( \AT{C} \vphi \big)$ is positive definite and hence the kernel of $\A{C}^{\phantom{\T}} \mb{C} \big( \AT{C} \vphi \big) \AT{C}$ is determined by the kernel of $\AT{C}$. As $\A{C}$ is an incidence (constant) matrix, we find that $\mb{Q}$ and $\mb{P}$ are also constant.
\end{remark}

\begin{remark}
    \label{re:dim_xtp}
    Again recalling \eqref{eq:md_mnaii} we observe
    \begin{align*}
        \mb{K}(\mb{x}) = \begin{bmatrix}
            \A{R}^{\phantom{\T}} \mb{G} \big( \AT{R} \vphi \big) \AT{R} & \A{L} & \A{V}\\
            -\AT{L} & &\\
            -\AT{V} & &
        \end{bmatrix}.
    \end{align*}
    Noting that $\mb{W} = \mb{Q}$ for MNA, due to the symmetry of $\mb{M}(\mb{x})$ in \autoref{re:dim_bf}, we find
    \begin{align*}
        \mbb{K}_\mb{Q}(\mb{x}) = \mb{Q}^\T \mb{K}(\mb{x}) \mb{Q} = \begin{bmatrix}
            \mb{Q}_\mr{C}^\T \A{R}^{\phantom{\T}} \mb{G} \big( \AT{R} \vphi \big) \AT{R} \mb{Q}_\mr{C} & \mb{Q}_\mr{C}^\T \A{V}\\
            -\AT{V} \mb{Q}_\mr{C} &
        \end{bmatrix}, \quad \mbb{K}_\mb{P}(\mb{x}) = \mb{Q}^\T \mb{K}(\mb{x}) \mb{P} = \begin{bmatrix}
            \mb{Q}_\mr{C}^\T \A{R}^{\phantom{\T}} \mb{G} \big( \AT{R} \vphi \big) \AT{R} \mb{P}_\mr{C} & \mb{Q}_\mr{C}^\T \A{L}\\
            -\AT{V} \mb{P}_\mr{C} &
        \end{bmatrix}.
    \end{align*}
    Determining $\mbb{W}(\mb{x})$ using $\mbb{K}_\mb{Q}(\mb{x})$ gives\cite{jansen2014}
    \begin{align*}
        \mbb{W} = \begin{bmatrix}
            \mb{Q}_\mr{V} \mb{Q}_\mr{R} &\\
            & \mb{W}_\mr{V}
        \end{bmatrix},
    \end{align*}
    where $\mr{im\, } \mb{Q}_\mr{V} = \ker \AT{V} \mb{Q}_\mr{C}$, $\mr{im\, } \mb{Q}_\mr{R} = \ker \AT{R} \mb{Q}_\mr{C} \mb{Q}_\mr{V}$ and $\mr{im\, } \mb{W}_\mr{V} = \ker \AT{V} \mb{Q}_\mr{C}$ are all constant such that $\mbb{W}$ is constant as well. This finally yields
    \begin{align*}
        \mbb{W}^\T \mbb{K}_\mb{P}(\mb{x}) = \begin{bmatrix}
            \mb{Q}_\mr{R}^\T \mb{Q}_\mr{V}^\T \mb{Q}_\mr{C}^\T \A{R}^{\phantom{\T}} \mb{G} \big( \AT{R} \vphi \big) \AT{R} \mb{P}_\mr{C} & \mb{Q}_\mr{R}^\T \mb{Q}_\mr{V}^\T \mb{Q}_\mr{C}^\T \A{L}\\
            -\mb{W}_\mr{V}^\T \AT{V} \mb{P}_\mr{C} &
        \end{bmatrix} = \begin{bmatrix}
            \mb{0} & \mb{Q}_\mr{R}^\T \mb{Q}_\mr{V}^\T \mb{Q}_\mr{C}^\T \A{L}\\
            -\mb{W}_\mr{V}^\T \AT{V} \mb{P}_\mr{C} &
        \end{bmatrix},
    \end{align*}
    thus the basis functions $\mbt{P}$ and $\mbt{Q}$ of $\mbb{W}^\T \mbb{K}_\mb{P}(\mb{x})$ are constant as well, and since $\mbb{W}^\T \mbb{f}(t)$ only depends on $t$, we find a unique solution for $\mbt{x}_\mb{P}$ that also only depends on $t$.
\end{remark}

\bibliography{biblio}

\clearpage

\section*{Author biography}
\begin{biography}{\includegraphics[width=66pt,height=86pt]{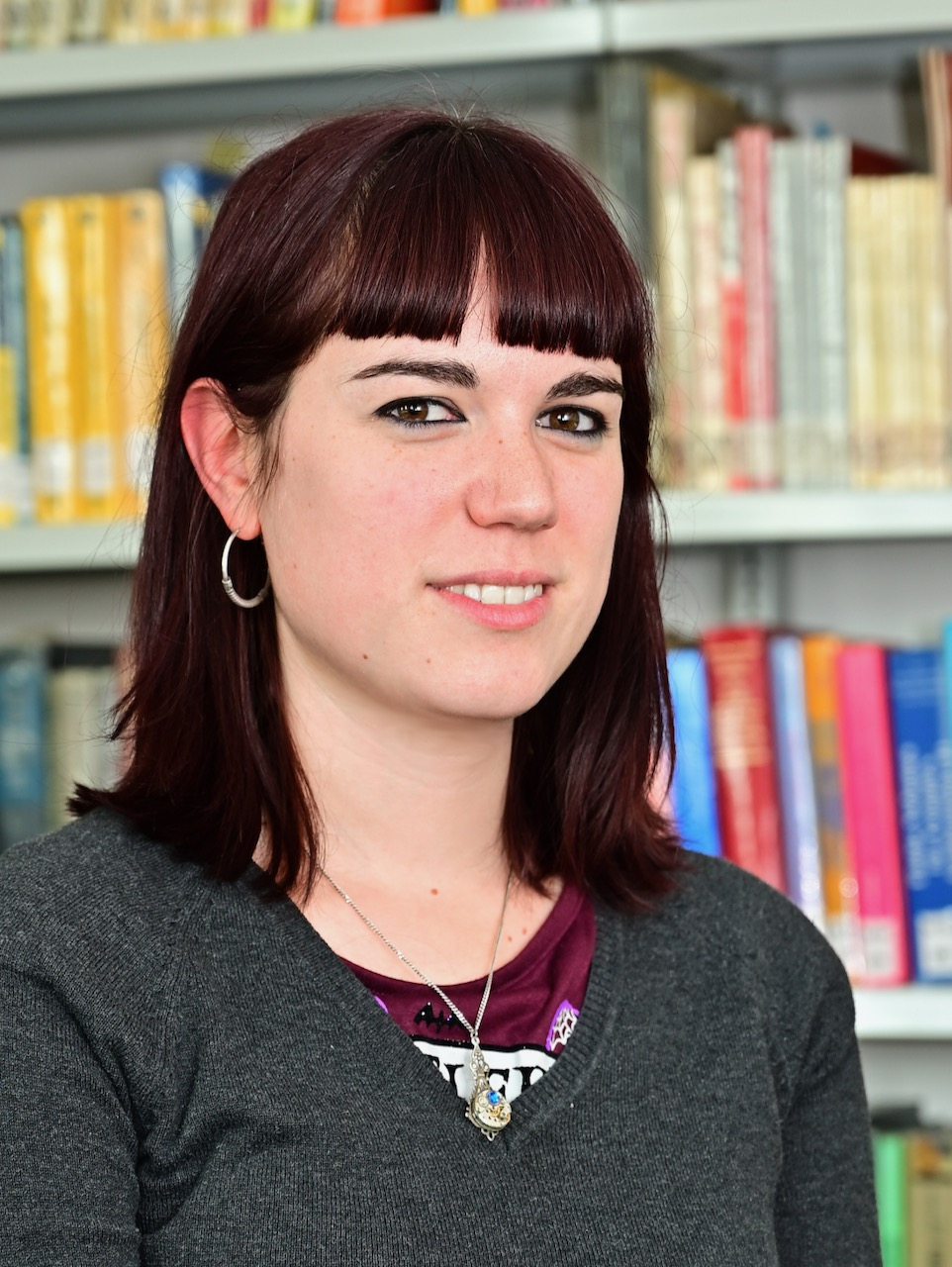}}{\textbf{Idoia Cortes Garcia} received her Bachelor in mathematics and Master in modelling for science and engineering from the Autonomous University of Barcelona. She obtained her Ph.D. in electrical engineering from the Technical University of Darmstadt in 2020. Since 2022 she is an assistant professor at the group of Dynamics and Control from the department of mechanical engineering at Eindhoven University of Technology. Her research interests include coupled multiphysical dynamical systems, differential algebraic equations, efficient time domain (co-)simulation methods and hybrid modelling approaches.}
\end{biography}
\vspace{0.5cm}
\begin{biography}{\includegraphics[width=66pt,height=86pt]{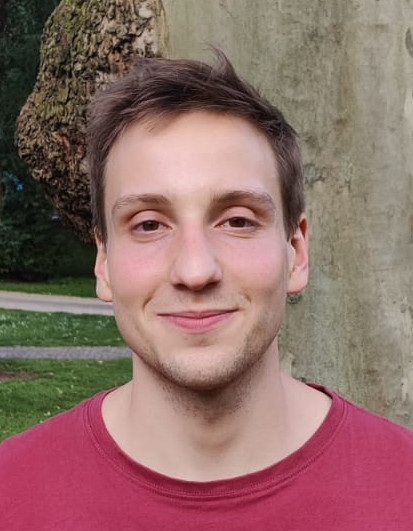}}{\textbf{Peter Förster} received his B.Sc. and M.Sc. degrees in electrical engineering from Technical University of Darmstadt. Currently, he is a doctoral researcher at the centre for analysis, scientific computing and applications at Eindhoven University of Technology and in the computational electromagnetics group at Technical University of Darmstadt. His research interests include circuit simulation, differential-algebraic equations, machine learning and digital twins.}
\end{biography}
\vspace{0.5cm}
\begin{biography}{\includegraphics[width=66pt,height=86pt]{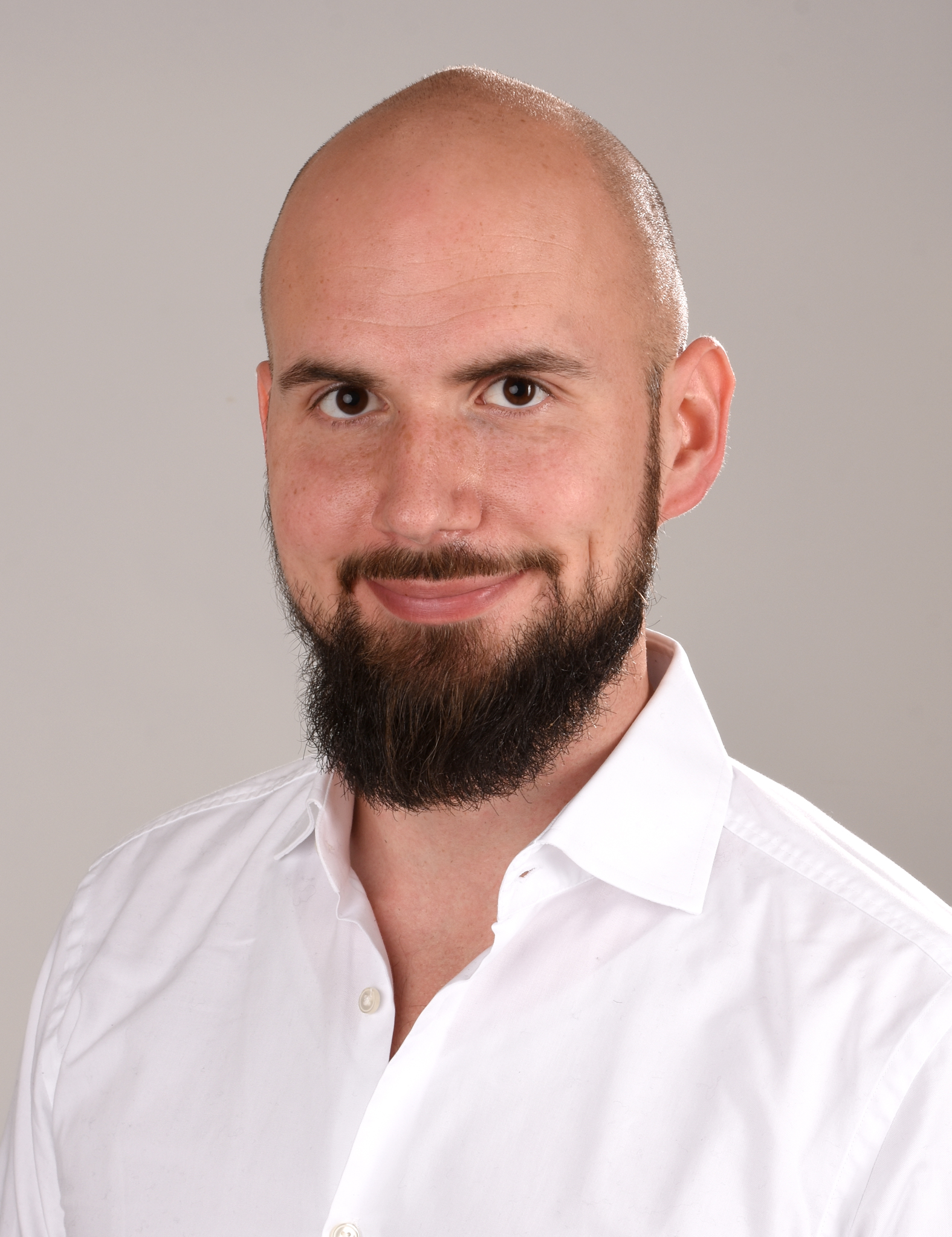}}{\textbf{Lennart Jansen} received his M.Sc. degree in mathematics from Universität zu Köln and his Ph.D. degree in mathematics from Universität zu Berlin. Afterwards, he held a post-doc position at Heinrich-Heine-Universität in Düsseldorf before starting work in the industry. One of the main research topics he worked on in the industry was AI-based recommendations for design improvement in the automobile industry. He switched to the photovoltaic industry in 2021 and currently works there on optimal control for combined electric-water-gas networks.}
\end{biography}
\vspace{0.5cm}
\begin{biography}{\includegraphics[width=66pt,height=86pt]{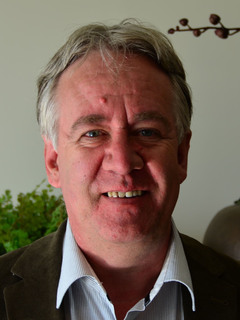}}{\textbf{Wil Schilders} received the M.Sc. degree in mathematics from Radboud University Nijmegen, Nijmegen, The Netherlands, in 1978, and the Ph.D. degree from the Trinity College Dublin, Dublin, Ireland, in 1980. He has been working in the electronics industry with Philips Research Laboratories, Eindhoven, The Netherlands, since 1980, and NXP, since 2006, where he developed algorithms for simulating semiconductor devices, electronic circuits, organic light emitting diodes, and electromagnetic problems (TV tubes, interconnects, and magnetic resonance imaging). Since 1999, he has been a Part-Time Professor of numerical mathematics for industry with the Technical University of Eindhoven, Eindhoven. He is currently the Managing Director of the Platform for Mathematics in The Netherlands. He is also the President of the European Consortium for Mathematics in Industry.}
\end{biography}
\vspace{0.5cm}
\begin{biography}{\includegraphics[width=66pt,height=86pt]{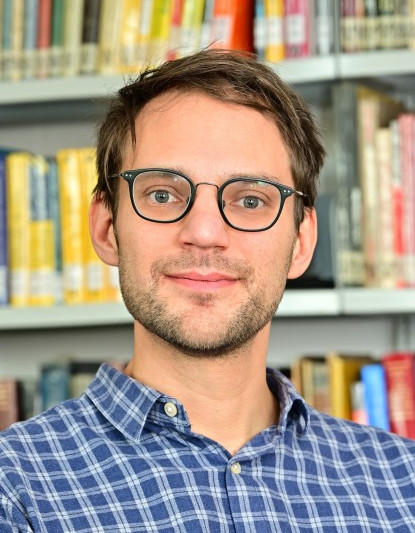}}{\textbf{Sebastian Schöps} received the M.Sc. degree in business mathematics and the joint Ph.D. degree from Bergische Universität Wuppertal and Katholieke Universiteit Leuven in mathematics and physics. He was appointed as a Professor of Computational Electromagnetics at Technische Universität Darmstadt within the interdisciplinary center of computational engineering, in 2012. His current research interests include coupled multi-physical problems, bridging computer aided design and simulation, parallel algorithms for high performance computing, digital twins, uncertainty quantification, and machine learning.}
\end{biography}

\end{document}